\documentclass[10pt,journal, compsoc]{IEEEtran}
\usepackage[dvipsnames]{xcolor}
\usepackage[nocompress]{cite} 
\usepackage[bookmarks,linkcolor=black]{hyperref} 
\usepackage{verbatim} 
\usepackage{ulem} 
\usepackage{makecell}
\usepackage{bm}
\usepackage{mathptmx}
\usepackage[pdftex]{graphicx}
\usepackage{epstopdf}
\usepackage{times}

\usepackage[ruled,vlined]{algorithm2e}  
\usepackage{amssymb,amsmath}
\usepackage{multirow}
\usepackage{caption}
\usepackage{paralist}
\usepackage{tikz}
\usepackage{multirow}
\usepackage{color}
\usepackage{ulem}
\usepackage{setspace}
\usepackage{hyphenat,balance}
\usepackage{dblfloatfix}
\usepackage{array}
\usepackage{url}
\usepackage{overpic}
\usepackage{contour}

\usepackage{amsfonts}
\usepackage{bbm}
\usepackage{diagbox}
\usepackage{url}
\usepackage{wrapfig}
\usepackage{hyphenat,balance,booktabs}

\usepackage{ragged2e}

\DeclareMathAlphabet{\pazocal}{OMS}{cmsy}{m}{n}

\def \etal {{\emph{et al}.}}

\def \etc {{\emph{etc}.}}
\def \eg {{\emph{e.g}.}}

\newcommand{\R}[1]{\textbf{R#1}}
\newcommand{\E}[1]{E{#1}}
\newcommand{\D}[1]{D{#1}}
\newcommand{\V}[1]{V$_{#1}$}
\renewcommand{\S}[1]{S$_{#1}$}
\newcommand{\SP}[1]{S$_{#1}'$}
\newcommand{\glyph}[2][1.3]{$\vcenter{\hbox{\includegraphics[height=#1\fontcharht\font`\B]{figures/glyph/#2.pdf}}}$}

\def \sys {{Reweighter}}

\usepackage{amssymb,amsmath}
\DeclareMathAlphabet{\mathcal}{OMS}{cmsy}{m}{n}
\usepackage{diagbox}

\newdimen\origiwspc
\newdimen\origiwstr

\ifCLASSOPTIONcompsoc
  \usepackage[nocompress]{cite}
\else
  \usepackage{cite}
\fi

\begin{document}

\title{Interactive Reweighting for\\ Mitigating Label Quality Issues}

\author{Weikai Yang, Yukai Guo, Jing Wu, Zheng Wang, Lan-Zhe Guo, Yu-Feng Li, and Shixia Liu
\IEEEcompsocitemizethanks{
\IEEEcompsocthanksitem W.~Yang, Y.~Guo, Z.~Wang, and S.~Liu are with the School of Software, BNRist, Tsinghua University. 
\IEEEcompsocthanksitem J.~Wu is with Cardiff University.
\IEEEcompsocthanksitem L.~Guo and Y.~Li are with Nanjing University.
}
}

\IEEEtitleabstractindextext{%
\justify
\begin{abstract}
Label quality issues, such as noisy labels and imbalanced class distributions, have negative effects on model performance.
Automatic reweighting methods identify problematic samples with label quality issues by recognizing their negative effects on validation samples and assigning lower weights to them.
However, these methods fail to achieve satisfactory performance when the validation samples are of low quality. 
To tackle this, we develop {\sys}, a visual analysis tool for sample reweighting.
The reweighting relationships between validation samples and training samples are modeled as a bipartite graph.
Based on this graph, a validation sample improvement method is developed to improve the quality of validation samples.
Since the automatic improvement may not always be perfect,
a co-cluster-based bipartite graph visualization is developed to illustrate the reweighting relationships and support the interactive adjustments to validation samples and reweighting results.
The adjustments are converted into the constraints of the validation sample improvement method to further improve validation samples.
We demonstrate the effectiveness of {\sys} in improving reweighting results through quantitative evaluation and two case studies.\looseness=-1
\end{abstract}

\begin{IEEEkeywords}
Training data quality, sample reweighting, bipartite graph visualization
\end{IEEEkeywords}}

\teaser{
\setcounter{figure}{0}
  \centering
  \vspace{-2mm}
  \includegraphics[width=\linewidth]{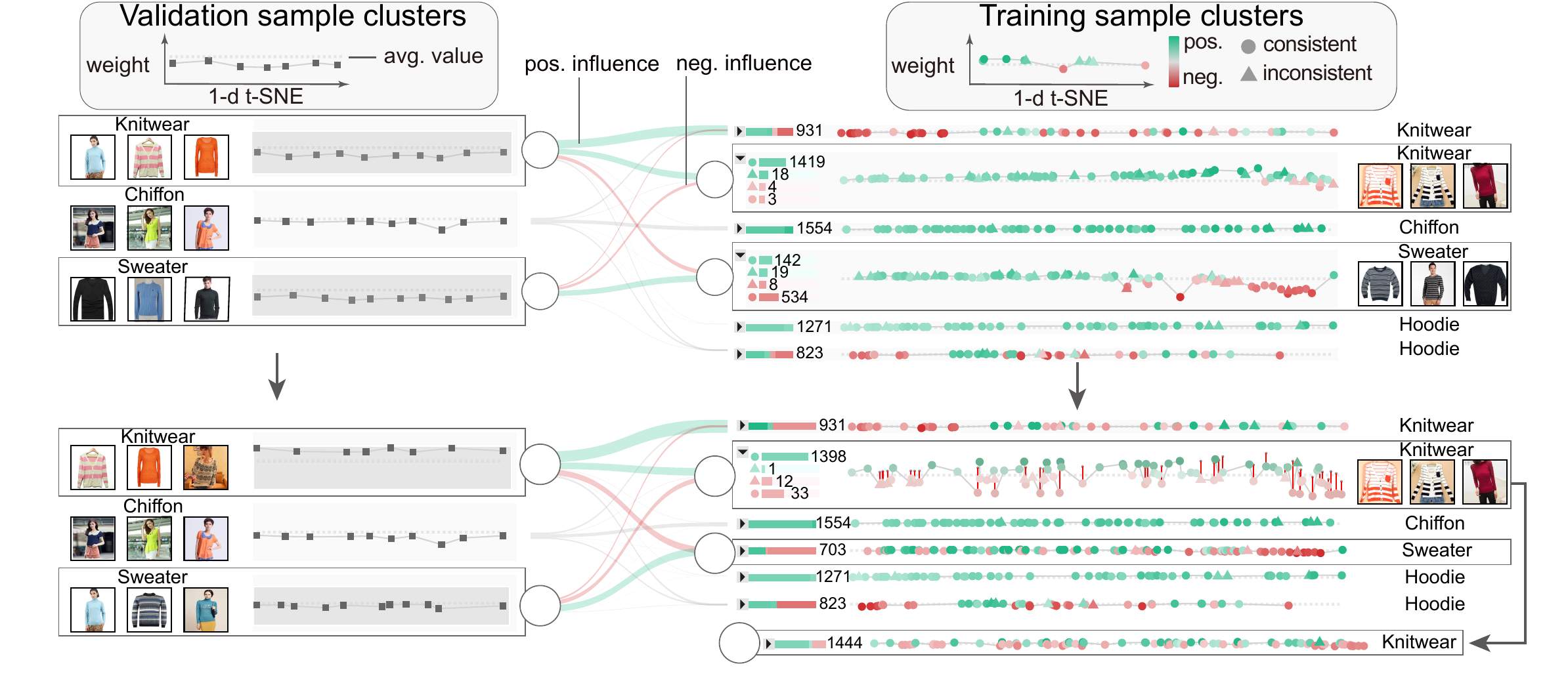}
  \put(-344,175){\fontsize{8}{8}\textrm{V${}_1$}}
  \put(-344,128.3){\fontsize{8}{8}\textrm{V${}_2$}}
  \put(-344,71) {\fontsize{8}{8}\textrm{V${}_1^\prime$}}
  \put(-344,24.5)  {\fontsize{8}{8}\textrm{V${}_2^\prime$}}
  \put(-286,166){\fontsize{8}{8}\textrm{S${}_1$}}
  \put(-286,133){\fontsize{8}{8}\textrm{S${}_2$}}
  \put(-286,68){\fontsize{8}{8}\textrm{S${}_1^\prime$}}
  \put(-286,42){\fontsize{8}{8}\textrm{S${}_2^\prime$}}
  \put(-278,11.5) {\fontsize{8}{8}\textrm{S${}_1^{\prime\prime}$}}
  \put(-338,97){\fontsize{8}{8}\textrm{(a) Initial results}}
  \put(-348,3){\fontsize{8}{8}\textrm{(b) After adjustments}}
  \vspace{-1mm}
  \captionsetup{margin=0.5cm}
  \caption{
  (a) The reweighting relationships between 3 (out of 14) validation sample clusters and 6 (out of 35) training sample clusters.
  \V1 and \V2 contain low-quality validation samples, resulting in many inconsistent training samples in \S1 and \S2.
  (b) After correcting the noisy labels of low-quality validation samples, increasing the weights of high-quality validation samples, and verifying inconsistent training samples, the reweighting results are improved ($\text{S}_1''$ and $\text{S}_2'$).
  } 
  \label{fig:teaser}
  \vspace{-6mm}
}
\maketitle
  \IEEEdisplaynontitleabstractindextext
\IEEEpeerreviewmaketitle
{
\fontsize{10}{10} 
\section{Introduction}
Addressing label quality issues, such as noisy labels and imbalanced class distributions is critical due to their pervasive presence in real-world applications.
For example, insufficient labeled data often compels model developers to resort to pseudo-labeling techniques for easier model training~\cite{zhang2021flexmatch}. 
However, this workaround comes at a steep cost by introducing label quality issues that can significantly downgrade model performance and reliability.
This casts a shadow on its applicability in real-world scenarios.
Identifying and mitigating these label quality issues becomes difficult and even impossible in the era of deep learning.
A more practical solution is to reduce the negative effects of problematic training samples with label quality issues on model training.
Validation-sample-based reweighting methods~\cite{ren2018learning, zhang2021learning} have achieved state-of-the-art performance in reducing such negative effects.
These methods first measure the model performance on a small set of samples, namely validation samples, which have clean labels and well represent training data.
Then a higher/lower weight is assigned to a training sample if it increases/decreases the model performance on the validation samples.
This forms the reweighting relationships between validation samples and training samples. 
Since the validation samples determine the reweighting results, it is crucial to ensure their quality for the performance of these reweighting methods~\cite{hoang2022maximising}.

High-quality validation samples should have clean labels and well represent training data~\cite{xiang2019interactive,yuan2021survey}. 
In response to this need, state-of-the-art reweighting methods automatically select training samples that are likely to have clean labels as validation samples based on their confidence metrics~\cite{zhang2021learning,xu2021faster}.
While this automatic selection offers advantages in terms of speed and reduced manual interference, it comes with two issues.
First, these automatically selected validation samples, despite being selected based on confidence metrics, can still contain noisy labels.
Second, an over-reliance on these confidence metrics might lead to the omission of samples that represent the diverse nature of the training data, thereby creating a skewed or biased representation.
This, in turn, can degrade the reweighting performance~\cite{hoang2022maximising}.
To ensure the cleanliness of the validation samples, model developers need to identify the validation samples with noisy labels.
To ensure their representativeness, they need to repetitively compare the distributions of validation and training samples to identify the less represented samples~\cite{song2022graph}.
This debugging process is labor-intensive and time-consuming~\cite{Zhang2022}.
Upon identifying the quality issues, model developers can make direct adjustments if they are familiar with the application domain and data. 
However, in situations where the domain is unfamiliar, collaboration with domain experts and/or crowd workers becomes essential~\cite{Angler2023}.
Such collaboration also requires a visual analysis tool to efficiently convey the identified quality issues and guide appropriate adjustments.
The challenging nature of the debugging process and the specific needs for the collaboration motivate us to develop {\sys}.
We model the reweighting relationships between validation samples and training samples as a bipartite graph.
Based on this graph, a validation sample improvement method is developed to improve the quality of validation samples.
Since the automatic improvement may not always be perfect, a co-cluster-based bipartite graph visualization is developed to illustrate the reweighting relationships. 
By offering clear insights into how the reweighting results are derived, this visualization facilitates more informed adjustments to the validation samples and reweighting results. 
These adjustments are converted into the constraints of the validation sample improvement method to improve the quality of validation samples.
This, in turn, leads to better reweighting results.
The aforementioned analysis process is repeated iteratively.
Upon satisfaction, the reweighting results are utilized for model training.

The effectiveness of the validation sample improvement method is demonstrated using three numerical experiments on four different datasets.
Two case studies are conducted to demonstrate the usefulness of {\sys} in iteratively improving the quality of validation samples and generating better reweighting results.
The main contributions of this work are:

\begin{compactitem}
\item\noindent{A visual analysis tool that helps generate better reweighting results by iteratively improving validation samples.}
\item\noindent{A method for automatically improving validation samples based on the reweighting relationships.}
\item\noindent{A bipartite graph visualization that illustrates the reweighting relationships and helps make informed adjustments.}
\end{compactitem}

\section{Related Work}
\subsection{Sample Reweighting}

Existing reweighting efforts can be classified into two categories~\cite{shu2019meta}: distribution-based methods and validation-sample-based methods. 

Distribution-based methods assign weights to training samples based on the data distribution. 
For example, Liu~\etal\cite{liu2015classification} first introduced the importance reweighting method into binary classification to tackle the challenges posed by noisy labels.
The weight of a sample is set as the probability of this sample being clean over the probability of it being noisy.
They also extended this method to multi-class classification~\cite{wang2017multiclass}.
Instead of estimating the data distribution, later work treated the sample weights as latent variables of a learning model, such as a Bayesian model~\cite{wang2017robust} or 
a deep learning model~\cite{liu2021label}, and then inferred the weights of samples by this model.
However, these methods require manually designing the reweighting function or the learning model.
This makes it difficult to fit different datasets.

To address this challenge, validation-sample-based methods assign a weight to each training sample with the goal of optimizing the model performance on validation samples.
These methods effectively reduce performance degradation caused by both noisy labels and imbalanced class distributions.
Ren~\etal\cite{ren2018learning} calculated the weights of the training samples using the loss gradients in the validation samples.
{A negative weight is assigned to a sample if its associated gradient is positive. 
Instead of calculating the weights from gradients, Meta-Weight-Net~\cite{shu2019meta} trains an extra model to learn the weights.
However, these methods require a pre-defined set of validation samples with clean labels and balanced class distributions.
To be more flexible, Zhang~\etal\cite{zhang2021learning} automatically selected training samples that are likely to have clean labels as validation samples. 
Although this method saves the efforts for pre-defining validation samples, 
the performance is downgraded if the selected samples contain noisy labels and/or fail to represent the
diversity of the training data~\cite{hoang2022maximising}.
Our method seeks to address these limitations by utilizing the reweighting relationships between validation samples and training samples. With these relationships and the corresponding visualization, experts can 1) easily identify and correct the noisy labels, and 2) identify the uncovered training samples and add some of them to increase the representativeness of these validation samples.

\subsection{Visual Quality Improvement of Training Data}

Many visualization methods have been proposed to improve the quality of training data~\cite{he2022where,chen2022towards}.
They are classified into two groups~\cite{whang2021data,yang2023foundation}: collecting more annotated data and correcting annotation noise.\looseness=-1

\noindent\textbf{Collecting more annotated data.}
A variety of methods have been proposed to improve the efficiency of the annotation process.
For example, Moehrmann~\etal\cite{moehrmann2011improving} employed a self-organizing map to allow the selection and simultaneous annotations of multiple similar images.
The strategy of placing similar samples closer together is then applied to annotate different types of data~\cite{kurzhals2016visual,khayat2019vassl,halter2019vian,eirich2021irvine}.
Active learning algorithms are also integrated to recommend informative samples for efficient annotation~\cite{junior2017analytic,dennig2019fdive,snyder2019interactive,sperrle2019viana,choi2021unified,jia2021towards,zhao2022human,yang2022diagnosing}.
Beyond efficiency, other methods seek to address the issue of dataset bias, which is introduced due to the distribution difference between training data and test data.
Chen~\etal\cite{chen2020oodanalyzer} developed OoDAnalyzer to visually detect Out-of-Distribution samples that are not covered by training data.
Yang~\etal\cite{yang2020drift} used the energy distance to measure the magnitude of change in data distribution when test data streams in. 
The idea of comparing the data distributions has also been adopted by many other visual analysis work~\cite{wang2020conceptexplorer,yeshchenko2021visual,gou2020vatld,he2022where}.
Different from these methods, our work focuses on correcting annotation noise in validation samples and making them well represent training data.

\begin{figure*}[!tb]
\centering
{\includegraphics[width=\linewidth]{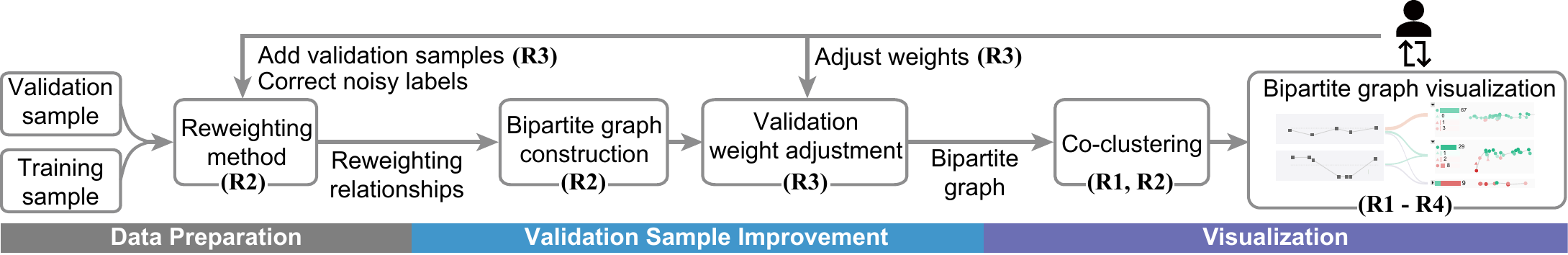}}
\caption{System overview.
The data preparation module extracts the reweighting relationships between validation samples and training samples.
The validation sample improvement module models the reweighting relationships as a bipartite graph and improves the quality of validation samples.
The visualization module facilitates interactive exploration and improvement on the validation samples.}
\vspace{-4mm}
\label{fig:overview}
\end{figure*}

\noindent\textbf{Correcting annotation noise in training data.}
Annotation noise is commonly present in real-world training data~\cite{yuan2021survey}.
Many methods have been proposed to correct them.
For training data with crowd information, crowd annotations and worker behavior are utilized to detect the annotation noise for correction~\cite{park2016c,liu2018crowsourcing,park2019cmed}.
For example, LabelInspect~\cite{liu2018crowsourcing} facilitates the verification of uncertain samples and unreliable workers based on crowd information analysis.
For training data without crowd information, the annotation noise is usually detected based on the difference between the sample annotations and their predictions made by a learning model~\cite{paiva2015approach,chen2021interactive,xiang2019interactive,bauerle2020classifier,chen2022towards,zhang2023labelvizier}. 
Among them, the most relevant one is DataDebugger~\cite{xiang2019interactive}, which develops a label propagation algorithm to identify and correct label noise based on the difference between their predictions and the expert-selected trusted items. 
Although this method improves the quality of training data, it has two issues. 
First, it requires to provide the exact label for each trusted item.
This takes more time since it needs the experts to examine the sample and even compare it with similar samples.
Second, it only handles the issue of label noise and does not work well for other label quality issues, such as imbalanced class distributions. 
Compared with this method, our work only requires the experts to indicate whether the label is clean or noisy, which demands less expertise and is more efficient. 
Moreover, our work handles not only the issue of label noise, but also the issue of imbalanced class distributions, which often occur in real-world training data.\looseness=-1

\section{Design of {\sys}}

\setcounter{figure}{3}
\begin{figure*}[b]
\centering
{
\vspace{-3mm}
\includegraphics[width=\linewidth]{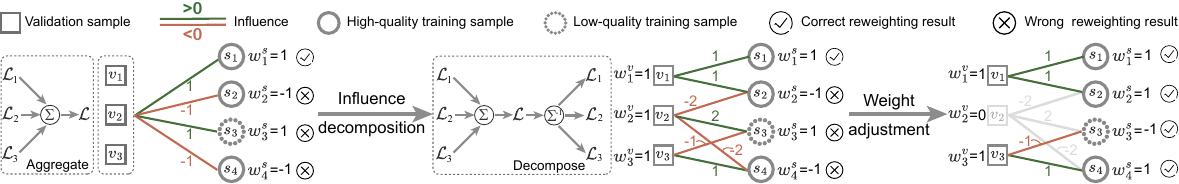}}
\put(-458,-6){(a)}
\put(-249,-6){(b)}
\put(-66, -6){(c) }
\caption{The benefit of introducing the bipartite graph in the validation sample improvement method: (a) without the bipartite graph, it is hard to diagnose why the reweighting results on $s_2\text{-}s_4$ are incorrect (high-quality samples $s_2$ and $s_4$ have negative weights; while low-quality sample $s_3$ have a positive weight); (b) 
with the bipartite graph, it is identified that $v_2$ leads to the incorrect results; (c) after reducing $w^v_2$ to zero, the reweighting results are improved.\looseness=-1}
\label{fig:toy-combined}
\end{figure*}

\subsection{Requirement Analysis}

We collaborated with four experts (\E1-\E4) to develop {\sys}.
\E1 is a software developer with five years of experience in improving the quality of training data.
\E2 and \E3 are two Ph.D. students with more than four years of experience in diagnosing quality issues in training data and improving model performance.
\E4 is a postdoc researcher with seven years of experience in few-shot learning, which also requires high-quality training data.
None of them are the co-authors of this work.
To understand how they improve the reweighting results in their work,
we interviewed each of them for about 45-60 minutes.
Based on the interviews and literature review, we summarized the following requirements for improving reweighting results.

\noindent\textbf{\R1. Examining validation samples and training samples}.
A previous study has shown that the quality of validation samples plays a critical role in the reweighting process~\cite{zhang2021learning}.
The experts also confirmed this and expressed the need to improve the validation samples. 
\E3 said that he would like to examine the quality of the validation samples and then check whether they positively influenced the reweighting results. 
In addition to examining the quality of validation samples, the experts mentioned that examining the training samples and the reweighting results are also useful, as they may reflect quality issues in validation samples.
\E1 said, ``{I would like to spend more time examining training samples with incorrect weights, such as low-quality samples with positive weights, to identify the quality issues in validation samples.}''
However, examining all the samples is time-consuming.
Thus, it is desirable to provide an informative overview of validation samples and training samples, and highlight the samples of interest.

\noindent\textbf{\R2. Understanding the reweighting relationships}.
After identifying the validation/training samples of interest, the experts need to understand the reweighting relationships related to these samples.
Without a comprehensive understanding of such relationships, it is difficult to make proper improvements and generate better reweighting results.
For example, \E3 commented that he would be more confident in adjusting validation samples after understanding how each validation sample influences the reweighting results.
\E1 added, ``{When I find a low-quality training sample with a positive weight, I would like to know which validation samples influence its weight and whether they also influence the weights of other training samples.''} 
However, it is difficult to examine all the relationships.
To simplify the analysis process, there is a need for a method that clusters the relationships for an overview and displays the details upon request.

\noindent\textbf{\R3. Improving the quality of validation samples}.
All the experts expressed the need to explicitly improve the quality of validation samples and hence generate better reweighting results through simple interactions.
For example, \E2 said that he would like to correct the labels of noisy validation samples because they would lead to incorrect reweighting results.
Meanwhile, the validation samples that positively influence the reweighting results should be strengthened.
When the validation samples cannot well represent training samples, he also wanted to add more representative validation samples for better coverage. 
Since the experts often found some incorrect reweighting results in their analysis, they also expressed the requirements for implicitly improving the quality of validation samples through adjustments to reweighting results.\looseness=-1

\noindent\textbf{\R4. Comparing the reweighting results before and after adjustments}.
Recent studies have shown that not all adjustments improve the quality of validation samples~\cite{ren2018learning,zhang2021learning}.
The experts also confirmed this.
For example, \E3 noted that while certain unsuitable adjustments might benefit the reweighting results of the training samples that he was analyzing, they could negatively impact other training samples.
Consequently, the experts highlighted the importance of comparing the reweighting results 
before and after the adjustments.
For example, \E4 commented, ``{The samples with large changes in their weights should be examined carefully.
If the adjustments result in a lot of unfavorable changes, I would like to reverse them and reanalyze the corresponding reweighting results.}''
Therefore, our tool is expected to emphasize  the differences before and after the adjustments to simplify the comparison process.\looseness=-1

\subsection{System Overview}
Motivated by the identified requirements, we develop {\sys} to support the interactive improvement of the reweighting results.
As shown in Fig.~\ref{fig:overview}, it contains three modules: data preparation, validation sample improvement, and visualization.

Given a set of validation samples and a set of training samples, the \textbf{data preparation module} initially extracts the reweighting relationships using a reweighting method (\R2). 
Theoretically, any validation-sample-based reweighting method can be directly used in {\sys}.
For our purpose, we employ the state-of-the-art reweighting method, Fast Sample Reweighting (FSR)~\cite{zhang2021learning}.
The \textbf{validation sample improvement module} models the extracted reweighting relationships between validation samples and training samples as a bipartite graph (\R2).
Based on the bipartite graph, the weights of validation samples are automatically adjusted for generating better reweighting results (\R3).
Next, the \textbf{visualization module} employs a co-clustering algorithm~\cite{Chakrabarti2004} to simultaneously group similar validation samples and training samples based on the constructed bipartite graph.
The co-clusters are visualized as a node-link diagram to simplify the exploration of the individual samples (\R1) and their reweighting relationships (\R2).
Using the interactive visualization, model developers can adjust the validation samples and the reweighting results (\R3).
The adjustments are then used to improve the quality of validation samples by updating the reweighting relationships in the data preparation module and adjusting the weights of the validation samples in the validation sample improvement module.
Furthermore, model developers can compare the reweighting results before and after the improvement to see if the adjustments are beneficial (\R4).

\section{Validation Sample Improvement}
\label{Sec:algorithm}
Since the quality of validation samples is critical in validation-sample-based methods, we introduce a method for improving validation samples. 
Common quality issues associated with validation samples include noisy labels and the lack of representative samples.
To address these issues, users usually correct the noisy labels and add the necessary samples.
The updated validation samples are then used to reweight the training samples.
During this process, a mutual influence between validation samples and training samples is observed. 
On the one hand, the quality of validation samples influences the reweighting results of training samples.
On the other hand, the reweighting results may reflect the quality issues of validation samples. 
Given this, we seek to capture and streamline the validation sample improvement process by modeling this influence as a bipartite graph.
With the bipartite graph, the weights of the validation samples can be improved based on their impact on the reweighting results.
This improvement process thus unfolds in two phases: 1) constructing a bipartite graph between validation samples and training samples; 2) adjusting the weights of validation samples based on this graph.

\setcounter{figure}{2}
\begin{figure}[t]
\centering
{
\includegraphics[width=\linewidth]{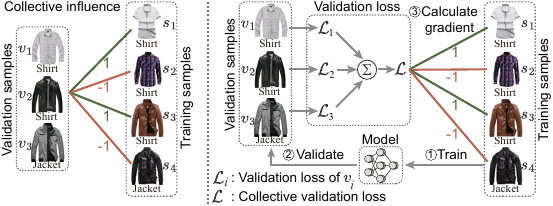}}
\caption{The collective influence on the reweighting results (left) is computed as the gradient of collective validation loss with regard to the weights of training samples (right).\looseness=-1}
\label{fig:influence}
\vspace{-3mm}
\end{figure}

\setcounter{figure}{4}
\begin{figure*}[!b]
\centering
{\includegraphics[width=\linewidth]{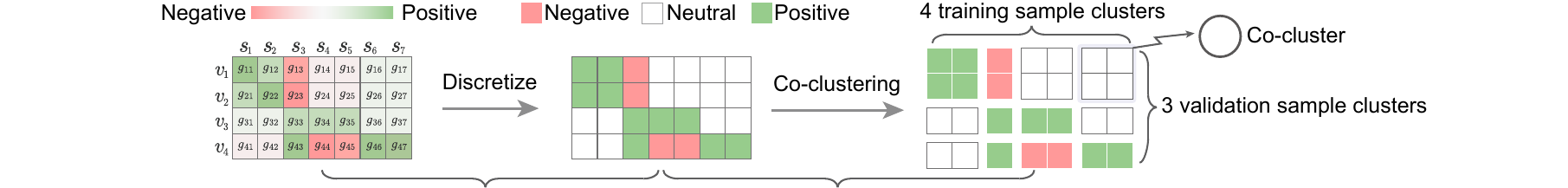}}
\put(-118.5,47){A}
\put(-363,-7.5){(a)}
\put(-246.7,-7.5){(b)}
\vspace{-1mm}
\caption{
The co-clustering pipeline: (a) the influence values are discretized into positive, neutral, and negative categories; (b) validation samples (rows) and training samples (columns) are grouped into clusters simultaneously.
}
\vspace{-2mm}
\label{fig:graph}
\end{figure*}

\subsection{Bipartite Graph Construction}
\label{subsec:graph_construction}
The state-of-the-art reweighting methods~\cite{ren2018learning,zhang2021learning} treat the validation samples as a whole when computing their influence on the weight of each training sample $s_j$.
As shown in Fig.~\ref{fig:influence}, the influence is computed as the gradient of the collective validation loss with regard to the weight of sample $s_j$.
However, mixing all validation samples together makes it difficult to identify the low-quality ones (Fig.~\ref{fig:toy-combined}(a)).
To address this issue, the collective influence from the validation sample set should be decomposed into the influence of individual samples (\R2).
The decomposition is achieved by building a bipartite graph between validation samples and training samples (Fig.~\ref{fig:toy-combined}(b)).\looseness=-1

The construction of the bipartite graph is fundamentally grounded on two key properties: 1) the validation loss is the sum of the loss on each validation sample; 2) the gradient operator is linear ($\nabla (f+g)=\nabla f+\nabla g$).
With these properties, the collective influence of the validation sample set on training samples can be decomposed into a set of influences $\{g_{ij}|1\le i\le m, 1\le j\le n\}$~\cite{Chatterjee2020Coherent}.
Here, $g_{ij}$ is the influence of the validation sample $v_i$ on the training sample $s_j$.
$m$ and $n$ are the numbers of validation samples and training samples, respectively.
The bipartite graph is then constructed by computing all $g_{ij}$s. 
With this graph, each validation sample $v_i$ is represented by a $n$-dimensional vector, $v_i=[g_{i1},g_{i2},\ldots,g_{in}]^\intercal$.
Each training sample $s_j$ is represented by a $m$-dimensional vector, $s_i=[g_{1j},g_{2j},\ldots,g_{mj}]^\intercal$.

\subsection{Validation Sample Weight Adjustment}
\label{subsec:weight-adjust}
The bipartite graph models how each validation sample influences the reweighting results.
Accordingly, the reweighting results can be improved by assigning lower/higher weights to the validation samples that have negative/positive effects on the reweighting results (\R3). 
For example, in Fig.~\ref{fig:toy-combined}(b), $s_2$ and $s_4$ are high-quality training samples with negative weights, and $s_3$ is a low-quality training sample with a positive weight.
Such reweighting results are incorrect.
Examining the reweighting relationships in the bipartite graph reveals that the validation sample $v_2$ leads to these incorrect results.
It incorrectly assigns negative weights to the high-quality training samples ($s_2, s_4$) and a positive weight to the low-quality training sample ($s_3$).
By reducing its weight to zero, the reweighting results are improved (Fig.~\ref{fig:toy-combined}(c)).
{From this example, it can be seen that the key to improving the reweighting results is to evaluate the quality of the validation samples and adjust their weights.
Next, we introduce how to evaluate the quality of validation samples by two measures: correctness and balancedness.\looseness=-1

\noindent\textbf{Correctness}.
A previous study has indicated that correct validation samples should assign positive weights to high-quality training samples and negative weights to low-quality training samples~\cite{xiang2019interactive}.
However, in practice, we do not know which training samples are of \textit{high quality} and which ones are of \textit{low quality}.
To solve this problem, we initially regard \textit{high-confidence} samples as high-quality and \textit{low-confidence} samples as low-quality.}
The confidence of a sample reflects how likely it is of high quality~\cite{zhang2021learning}.
Then we assess the correctness by examining whether the validation samples generate correct reweighting results on high-quality samples and low-quality samples.
This can be regarded as a binary classification problem.
We thus employ the binary cross-entropy, a commonly used loss for binary classification~\cite{goodfellow2016deep}, to measure the deviation from the correct reweighting results:

\begin{equation}
\mathcal{L}_{\text{c}}(w^s_1,\ldots,w^s_n) = \sum_{s_j\in S_{+}} -\log\phi(w^s_j)+\sum_{s_j\in S_{-}}-\log(1-\phi(w^s_j)),
\label{eq:loss_bce}
\end{equation}
where $w_j^s$ is the weight of training sample $s_j$, $S_{+}$ and $S_{-}$ are the sets of high-quality and low-quality samples, respectively.
$\phi(w)=1/(1+e^{-w})$ is a sigmoid function to normalize the weight to $(0, 1)$.
When minimizing the loss defined in Eq.~(\ref{eq:loss_bce}), the first term encourages positive weights on high-quality samples, while the second term encourages negative weights on low-quality samples. 

\begin{figure*}[!b]
\centering
{\includegraphics[width=\linewidth]{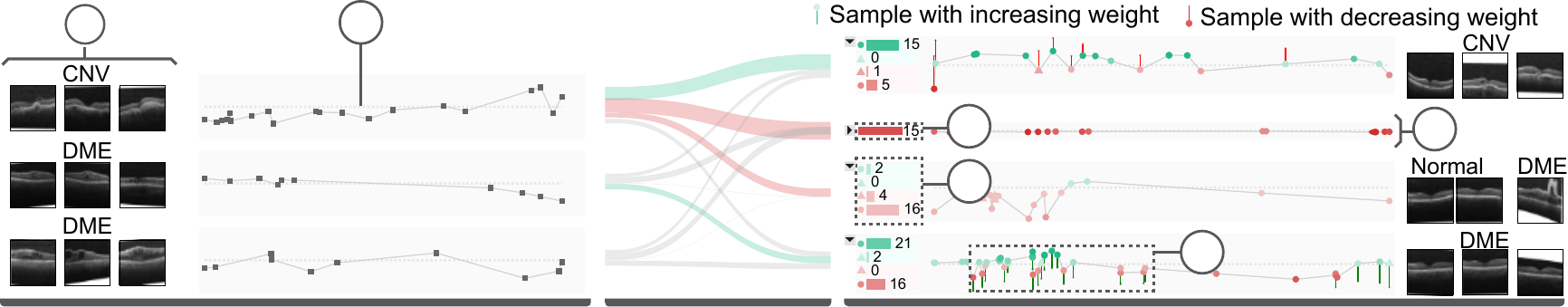}}
\put(-494.6,90.6){A$_{1}$}
\put(-404,90.6){A$_{2}$}
\put(-202.7,55.5){C$_{2}$}
\put(-202.7,38){C$_{1}$}
\put(-125.5,14.5){C$_{4}$}
\put(-49,55.5){C$_{3}$}
\put(-482,-9.5){(a) Validation sample cluster (node)}
\put(-315,-9.5){(b) Influence (link)}
\put(-180,-9.5){(c) Training sample cluster (node)}
\vspace{-1mm}
\caption{The visual design of the cluster view.}
\label{fig:vis}
\end{figure*}

\noindent\textbf{Balancedness}.
According to the study of He~\etal\cite{he2009learning}, balanced validation samples assign equal weights to high-quality training samples across different classes. 
This indicates that the weight distribution over classes is uniform.
In our implementation, the weight distribution is computed on high-quality samples. 
Here, only high-quality samples are considered because their labels are more reliable.
Let $C$ be the number of classes and $S_{c+}$ be the set of high-quality samples of class $c$ ($1\le c\le C)$.
The weight distribution is obtained by summing the sample weights in each class $c$ and then dividing by the total weights: $p_c = (\sum_{s_j\in {S_{c+}}}w^s_j)/(\sum_{s_j\in {S_{+}}}w^s_j)$ ($1\le c\le C)$.
Accordingly, the balancedness is measured by 
the deviation of the weight distribution over all classes from the uniform distribution.
It is computed by the Shannon entropy loss~\cite{gray2011entropy}:
\begin{equation}
\mathcal{L}_{\text{b}}(w^s_j,\ldots,w^s_j) = \sum_{c=1}^{C}p_c\log p_c.
\label{eq:loss_balance}
\end{equation}
A smaller entropy loss reflects a more balanced distribution.

With these two quality measures, the validation sample weight adjustment is formulated as a multi-task learning problem, which aims to minimize the combination of the two corresponding measures.
However, it is prohibitively expensive and difficult to manually tune the optimal weighting parameters for combining them.
To tackle this issue, we adopted multi-task learning~\cite{kendall2018multi} to automatically adjust the weighting parameters.
The basic idea is to reduce the contribution of the measure with higher uncertainty.
Let $\sigma_c$ and $\sigma_b$ denote the uncertainty of $\mathcal{L}_c$ and $\mathcal{L}_b$, respectively, the optimization goal is defined as:

\begin{align}
\underset{w^v_i,\sigma_c,\sigma_b}{\mathrm{minimize}}\quad \frac{1}{\sigma_c^2}\mathcal{L}_{\text{c}}&(w^s_1,\ldots,w^s_n)+\frac{1}{\sigma_b^2}\mathcal{L}_{\text{b}}(w^s_1,\ldots,w^s_n)+\log\sigma_c\sigma_b\nonumber\\
\quad\mathrm{s.t.}\quad &w^s_j=\sum_{i=1}^{m}w^v_i g_{ij},\ \forall j; \qquad w^v_i\ge 0,\ \forall i, 
\label{eq:opt}
\end{align}
where $w_i^v$ is the weight of validation sample $v_i$.
The first term and the second term correspond to the correctness and the balancedness, respectively.
The last term penalizes too large settings of $\sigma_c$ or $\sigma_b$.
The values of $\sigma_c$ and $\sigma_b$ are learned during the optimization.

To ensure the non-negative constraints ($w_i^v\ge 0$) in the optimization process, we employ Projected Gradient Descent~\cite{boyd2004convex} to solve this optimization problem.
The basic idea is to compute the gradient and project it to the subspace tangent to the constraints.
This ensures the projected gradient satisfies the constraints.
\section{Visualization}
\label{Sec:vis}

Better understanding the reweighting relationships between validation samples and training samples facilitates the adjustments to the validation samples and reweighting results.
To effectively reveal a large number of such relationships, we first co-cluster the validation samples and training samples based on the constructed bipartite graph.
Then a bipartite graph visualization is developed to visually illustrate the reweighting relationships.

\subsection{Co-Clustering}
The bipartite graph usually contains thousands of samples and hundreds of thousands of reweighting relationships.
Displaying all of them will cause severe visual clutter.
To facilitate the exploration, we group similar validation samples and similar training samples simultaneously through co-clustering.
The most widely used co-clustering algorithm is Spectral Co-clustering~\cite{dhillon2001co,chen2022towards}.
A drawback of this algorithm is that it involves costly matrix decomposition whose 
time complexity is $O((n+m)^3)$. 
Here, $m$ and $n$ are the numbers of validation samples and training samples, respectively.
Thus, it is intractable for a large graph. 
A method to solve this problem is Fully Automatic Cross-Associations (FACA)~\cite{chakrabarti2012graph}, which is an information-theory-based method.
This method can reduce the time complexity to $O(nm)$.
However, to use this method, the influence values ($g_{ij}$) need to be discrete. 
In our case, although the influence values are continuous, users are more interested in the non-zero influence values because they affect the reweighting results mostly.
Therefore, we discretize the continuous values into three categories: positive, neutral, and negative (Fig.~\ref{fig:graph}(a)).
To decide the threshold $\varepsilon$ among the positive ($\ge\varepsilon$), neutral (between $-\varepsilon$ and $\varepsilon$), and negative categories ($\le-\varepsilon$),
we have experimented with four datasets. 
The experimental results show that $\varepsilon=0.05$ is the best value (see supplemental material for more details).

With the discretization,  
we then employ the FACA method~\cite{chakrabarti2012graph} to build the co-clusters (Fig.~\ref{fig:graph}(b)).
A co-cluster consists of a pair of highly relevant validation sample cluster and training sample cluster (Fig.~\ref{fig:graph}A).
The basic idea of FACA is to greedily group samples into clusters while maintaining the high purity of each co-cluster.
The purity is defined as the fraction of influence values belonging to the dominant influence category within each co-cluster.
Take Fig.~\ref{fig:graph} as an example, 
the influence category of each co-cluster only belongs to one category (positive, neutral, or negative), and is therefore pure.
In addition, FACA automatically determines the optimal number of clusters based on the Minimum Description Length principle~\cite{rissanen1978modeling}.

In practice, the number of validation samples is typically small, often in the hundreds.
In contrast, the number of training samples is much larger, reaching tens of thousands or even more. 
To better convey these training samples, we hierarchically cluster them by using the hierarchical clustering method developed by Chen \etal~\cite{chen2022towards}.
Specifically, we fix the validation sample clusters and recursively apply FACA to divide each high-level training sample cluster into sub-clusters.\looseness=-1

\subsection{Bipartite Graph Visualization}
The visualization consists of two views: 1) a cluster view to provide an overview of reweighting relationships and help select the samples of interest; 2) a sample view to reveal the critical training samples that are influenced by the selected validation samples or the critical validation samples that influence the selected training samples.\looseness=-1

\subsubsection{Cluster View}
A node-link diagram is employed in the cluster view to visualize the bipartite graph based on the co-clustering result (Fig.~\ref{fig:vis}).
In the diagram, each node represents a validation sample cluster (Fig.~\ref{fig:vis}(a)) or a training sample cluster (Fig.~\ref{fig:vis}(c)), and
each link represents the influence between a validation sample cluster and a training sample cluster (Fig.~\ref{fig:vis}(b)).
We choose the node-link diagram because of its intuitiveness for understanding the reweighting relationships~\cite{ghoniem2005readability}.\looseness=-1

\noindent\textbf{Validation sample cluster} (\R1).
In each cluster, a dark gray square \glyph[1.0]{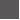} represents a validation sample.
The $y$-position of a square encodes the weight ($w^v_i$) of the validation sample.
A dotted line in the middle of each cluster (Fig.~\ref{fig:vis}A${}_2$) represents the average weight of all validation samples.
Squares above/below the line are the samples with larger/smaller weights.
Along the $x$-axis, placing similar samples together facilitates faster identification of related samples by examining the neighbors of a known one, thus streamlining the exploration process. 
In light of this, t-SNE is employed to project validation samples onto a one-dimensional space because of its effectiveness in preserving the neighborhood relationships between samples~\cite{li2022unified}.
As the number of samples in each cluster can vary widely, the perplexity in the t-SNE projection is adaptively set as the square root of the number of samples in the cluster, which gives satisfactory results in practice~\cite{robinson2020tree}.
Other hyperparameters used in t-SNE are fixed with the default values in Scikit-learn.
To better convey the trend of the weight change, the validation samples are connected by a polyline along the $x$-axis~\cite{saket2018task}.
To help users quickly understand the content of each cluster, we sample representative images and display them on the left side of the cluster (Fig.~\ref{fig:vis}A${}_1$).

\noindent\textbf{Training sample cluster} (\R1).
The training sample cluster employs a similar design to that of the validation cluster.
Here we describe the differences.
In each cluster, the dotted line in the middle is the line of weight zero.
The samples positioned above the line have positive weights, whereas those below it bear negative weights.
To enhance the clarity of this distinction, we have employed a double-encoding method that combines vertical positioning with a diverging color scheme.
In this scheme, \textcolor[rgb]{0.15,0.7,0.5}{green} represents positive weights, and \textcolor[rgb]{0.78,0.18,0.18}{red} represents negative weights. 
Moreover, the lightness of the color represents the absolute value of the weight $\lvert w\rvert$.
Darker colors indicate larger absolute values.
Each sample is represented by a circle or a triangle glyph to indicate whether its weight is consistent with the associated confidence value (\glyph[1]{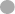}) or not (\glyph[1]{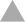}). 
For example, an inconsistent sample with a high-confidence value but a negative weight is denoted as \glyph[1]{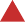}. 
To facilitate the examination of inconsistent training samples, a bar chart (Fig.~\ref{fig:vis}C${}_1$) is used to show the number of samples of different types (\glyph[1]{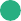}, \glyph[1]{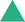}, \glyph[1]{redtri}, and \glyph[1]{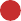}).
By default, the clusters with fewer inconsistent samples are collapsed (Fig.~\ref{fig:vis}C${}_3$) by stacking the bars horizontally and removing the sample images (Fig.~\ref{fig:vis}C${}_2$).\looseness=-1

When a cluster contains too many samples, we select a subset of representative ones, giving priority to high-quality samples $S_+$ and low-quality samples $S_-$ due to their importance in evaluating the quality of validation samples.
Motivated by the outlier-biased sampling method~\cite{xiang2019interactive,yuan2020evaluation}, we try to preserve the sample distribution while prioritizing the sampling of training samples in $S_+$ and $S_-$.
Specifically, the sampling probability of a sample $x$ is proportional to $1/\rho(x)+1[x\in S_+\cup S_-]$, where $\rho(x)$ is the density of its local region, and $1[x\in S_+\cup S_-]$ indicates whether it is a high-quality sample or a low-quality sample.\looseness=-1

\noindent\textbf{Links} (\R2).
A link between a validation sample cluster and a training sample cluster is represented by a line strip.
The width of the strip encodes the sum of the influence values between the associated validation samples and training samples.
Green strips and red strips represent positive and negative influence values, respectively.
To reduce visual clutter, we first minimize the strip crossings by reordering the validation sample clusters and the training sample clusters. 
The ordering is initialized by the widely-used barycenter heuristic~\cite{sugiyama1981methods} and then improved by swapping adjacent clusters to further reduce the number of crossings~\cite{gansner1993technique}.
Next, the strips with small influence values are shown with light gray and serve as context.
They will be highlighted when hovering over the associated clusters.

\subsubsection{Sample View}
After selecting the samples of interest in the cluster view, users can examine the associated validation/training samples and their detailed reweighting relationships in the sample view (\R1, \R2).
We adopt the adjacency-list design because it can compactly represent the reweighting relationships between the training samples and validation samples and their content~\cite{hlawatsch2014visual}.
When the exploration starts by \textbf{selecting training samples}, each row corresponds to a selected training sample.
As shown in Fig.~\ref{fig:sample}A, its image content is shown at the beginning of the row, followed by the top-3 positively contributing validation samples and the top-3 negatively contributing validation samples.
For each training sample $s_j$, its weight ($w^s_j$) is placed under the image.
For each contributing validation sample $v_i$, its weighted influence value ($w^v_i g_{ij}$) is placed to directly reveal how much $v_i$ contributes to $w^s_j$.
The circle/triangle glyphs representing the sample types (e.g., an inconsistent sample \glyph[1]{greentri}/\glyph[1]{redtri}) are displayed beside the weights/values.
When the exploration starts by \textbf{selecting validation samples}, each row corresponds to a selected validation sample (Fig.~\ref{fig:sample}B).
For each validation sample $v_i$, its weight ($w^v_i$) is placed under the image content.
The top-3 training samples that are positively/negatively influenced by it and their associated influence values are displayed in the corresponding row.\looseness=-1

\begin{figure}[!tb]
\centering
{\includegraphics[width=\linewidth]{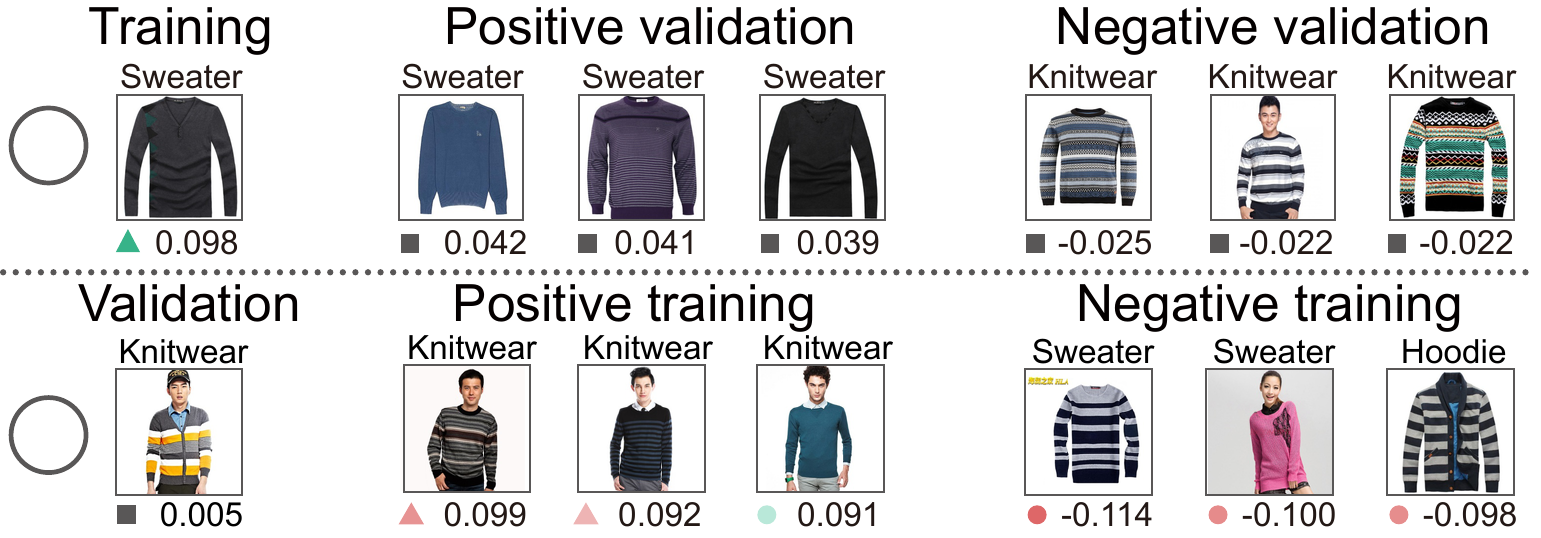}}
\put(-248.5,61){A}
\put(-248,13){B}
\vspace{-2mm}
\caption{The visual design of the sample view.}

\vspace{-5mm}
\label{fig:sample}
\end{figure}

\subsection{Interactive Adjustment and Comparison}
{\sys} allows users to improve the quality of validation samples by interactively adjusting validation samples and training samples.
After an adjustment, the users can also compare the reweighting results before and after the adjustment.\looseness=-1

\noindent\textbf{Interactive adjustment} (\R3). 
{\sys} supports the adjustments to both validation samples and training samples.

\textit{Adjusting the validation samples}.
{\sys} provides several ways to adjust the validation samples.
If the users find a mislabeled validation sample, they can right-click the sample and relabel (\glyph{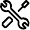}) it in the pop-up menu.
If the users identify some training samples that are not covered by the validation samples, they can add (\glyph{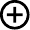}) them to the validation samples to improve the coverage.
The influence values in the reweighting relationships will be updated accordingly after the two adjustments.
In addition, the users can adjust the weights of validation samples if they are not appropriate.
Since it is difficult for the users to specify the exact weights, we allow them to simply indicate the direction of the weight adjustment by dragging the samples up (increase) or down (decrease). 
The adjustment is formulated as the inequality constraint of the weight adjustment optimization problem defined in Eq.~(\ref{eq:opt}).
In our implementation, the inequality constraint is set as $w^v_i\ge (1+\gamma) \tilde{w}^v_i$ when increasing the weight and as $w^v_i\le (1-\gamma) \tilde{w}^v_i$ when decreasing the weight. 
Here $\tilde{w}^v_i$ is the previous weight, and $\gamma=0.1$.
The details regarding this default value can be found in supplemental material.

\textit{Adjusting the training samples.}
As the quality measures of validation samples are calculated based on the high-quality sample set $S_+$ and the low-quality sample set $S_-$, users can implicitly improve the quality of validation samples by refining $S_+$ and $S_-$.
To this end, users first select a set of samples of interest by brushing them in a training sample cluster or clicking a bar in a bar chart associated with a cluster. 
Then they can right-click the selected samples and verify them as high-quality (\glyph{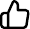}) or low-quality (\glyph{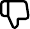}) in the pop-up menu. 
In addition, they can verify samples as high-quality by increasing their weights or as low-quality by decreasing their weights.
The user-verified high-quality and low-quality samples are added to $S_+$ and $S_-$, respectively.
With the refined $S_+$ and $S_-$, the calculation of the correctness (Eq.~(\ref{eq:loss_bce})) and balancedness (Eq.~(\ref{eq:loss_balance})) are updated accordingly, which affects the optimization results in Eq.~(\ref{eq:opt}).

\noindent\textbf{Comparison of the reweighting results} (\R4).
After the users finish their adjustments, the reweighting results are updated accordingly by re-solving the optimization problem defined in Eq.~(\ref{eq:opt}).
To facilitate the result comparison between before and after the adjustments, the users can switch to the ``diff'' mode, which highlights significant changes.
In this mode, the samples are positioned based on their current reweighting results.
For samples with significant weight changes, each has a green or red line connecting its previous and current positions to highlight the change (Fig.~\ref{fig:vis}C${}_4$).
The color indicates whether the weight is \textcolor[rgb]{0.15,0.7,0.5}{increased} or \textcolor[rgb]{0.78,0.18,0.18}{decreased}.
By default, the samples with weight changes among the top 10\% are considered to have significant changes, and the percentage (10\%) can be modified by the users.
After examining the weight changes, they can undo the adjustments if the changes are unsatisfactory.

\section{Evaluation}

We evaluated our validation sample improvement method using three separate experiments.
These experiments were conducted automatically, requiring no human intervention.
Moreover, through two case studies, we demonstrate how {\sys} enables users to interactively improve validation samples, leading to better reweighting results.\looseness=-1

\subsection{Quantitative Evaluation}
\label{subsec:quantitative}
This section evaluates the effectiveness of the improved validation samples in the separate scenarios of noisy labels and imbalanced class distributions, as well as the scenario of their combination.

\noindent\textbf{Datasets}.
Four datasets are used in the evaluation.
The first two, \textbf{CIFAR10} and \textbf{CIFAR100}, are commonly used benchmark datasets for evaluating sample reweighting methods~\cite{ren2018learning, zhang2021learning}.
CIFAR10 contains 60,000 samples in 10 classes, with 
5,000 training samples and 1,000 test samples per class.
Similarly, CIFAR100 contains 60,000 samples in 100 classes, with 500 training samples and 100 test samples per class.
These two datasets are clean and balanced.
We thus simulated noisy labels and imbalanced class distributions for these two datasets.
The third one, \textbf{Clothing} dataset~\cite{xiang2019interactive}, contains 37,497 samples in 14 different classes (T-shirt, Shirt, Knitwear, \etc).
We used 36,000 samples for training and 1,497 samples for testing.
As the sample labels are automatically extracted from the associated text descriptions, this dataset is noisy with 38.3\% of the samples being mislabeled. 
The fourth one, \textbf{OCT} (Optical Coherence Tomograph) dataset~\cite{kermany2018identifying}, contains 20,000 retinal OCT images (17,000 for training and 3,000 for test).
The images are of four classes: Normal, Choroidal NeoVascularization (CNV), DiabeticMacular Edema (DME), and Drusen.
This dataset is imbalanced as there are only 1,000 training samples of DME and 1,000 of Drusen, compared to 10,000 of Normal and 5,000 of CNV.
The imbalance factor, which is the ratio between the number of training samples in the largest class and that in the smallest class, is $10000/1000=10$.

\begin{table}[!b]
\centering
\caption{The noise ratios of the simulated noisy labels.}
\setlength\tabcolsep{14pt}
\centering
\begin{tabular}{ccccc}
\toprule
\multirow{2}{*}{Dataset} &  \multicolumn{4}{c}{\# labeled samples per class}  \\
                         &   10 & 20 & 50 & 100  \\
\midrule
{CIFAR10} & 0.48 & 0.42 & 0.36 & 0.32   \\
\midrule
{CIFAR100} & 0.70 & 0.65 & 0.55 & 0.45   \\
\bottomrule
\end{tabular}
\label{tab:ratio}
\end{table}

\noindent\textbf{Experimental settings}.
To evaluate the effectiveness of the developed validation sample improvement method, we combined it with a state-of-the-art reweighting method, FSR~\cite{zhang2021learning}.
We fed the improved validation samples into FSR and evaluated whether the generated weights further improved model performance.
For comparison, we used Uniform (no reweighting) and FSR~\cite{zhang2021learning} as the baseline methods.
Following the settings of FSR, we utilized WideResNet-28-10~\cite{zagoruyko2016wide} for low-resolution images (CIFAR10 and CIFAR100) and ResNet50~\cite{he2016deep} for high-resolution images (Clothing and OCT).
The models were also trained for 255 epochs with a cosine learning rate decay.
The numbers of validation samples are 100, 200, 140, and 80 for CIFAR10, CIFAR100, Clothing, and OCT datasets, respectively.
For evaluation in the scenario of noisy labels, the Clothing dataset with real-world label noise is used. 
In addition, CIFAR10 and CIFAR100 with simulated noisy labels are used to evaluate the classification accuracy under different noise ratios.
Pseudo labeling utilizes a trained model to generate labels for unlabeled samples and then incorporates these newly labeled samples back into the training set to refine the model further.
This method closely simulates human-like labeling errors~\cite{gu2023instance}.
Using this method, we first randomly selected and labeled $k\in\{10,20,50,100\}$ samples for each class in CIFAR10 and CIFAR100 and built four classification models~\cite{yang2022diagnosing}.
The pseudo labels of the remaining samples in the two datasets are then derived using these classification models.
The noise ratios of the pseudo labels are summarized in Table~\ref{tab:ratio}.
For evaluation in the scenario of imbalanced class distributions, the OCT dataset is used.
Similarly, we used CIFAR10 and CIFAR100 with simulated imbalanced class distributions to evaluate the accuracy under different imbalance factors.
The simulation was performed by sampling a subset of training samples following the Pareto distribution with imbalance factors $\lambda\in\{5, 10, 20, 50\}$.
In the combined scenario, we simulated label noise using different numbers of labeled samples per class ($k$) in the OCT dataset, and simulated imbalanced class distributions with different imbalance factors ($\lambda$) on the Clothing dataset.
For CIFAR10 and CIFAR100, we evaluated the accuracy under all the combinations of $k$ and $\lambda$.
We repeated each experiment three times and reported the average accuracy on the test set.

\begin{table}[!t]
\centering
\caption{Test set accuracy under noisy labels.}
\setlength\tabcolsep{6pt}
\centering
\begin{tabular}{cccccc}
\toprule
\multirow{2}{*}{Dataset} & \multirow{2}{*}{Method} & \multicolumn{4}{c}{\# labeled samples per class}  \\
                         &                         & 10 & 20 & 50 & 100  \\
\midrule
\multirow{3}{*}{CIFAR10} 
& Uniform  & 56.3\% & 62.0\% & 67.1\% & 70.5\%  \\
& FSR  & 67.6\% & 74.7\% & 78.6\% & 81.2\%  \\
& Ours  & \textbf{68.4\%} & \textbf{75.4\%} & \textbf{79.0\%} & \textbf{81.4\%}  \\
\midrule
\multirow{3}{*}{CIFAR100} 
& Uniform  & 30.4\% & 35.9\% & 43.9\% & 50.4\%  \\
& FSR  & 37.0\% & 43.5\% & 53.4\% & 58.8\%   \\
& Ours & \textbf{38.1\%} & \textbf{44.2\%} & \textbf{53.7\%} & \textbf{59.1\%}  \\
\bottomrule
\end{tabular}
\\
\vspace{1mm}
(a) Simulated noisy labels.
\vspace{2mm}

\centering
\setlength\tabcolsep{13pt}
\begin{tabular}{cccc}
\toprule
Dataset & Noise ratio & Method & Accuracy \\
\midrule
\multirow{3}{*}{Clothing} & \multirow{3}{*}{0.383}
& Uniform & 57.8\%    \\
& & FSR & 70.5\%    \\
& & Ours & \textbf{71.9\%}    \\

\bottomrule
\end{tabular}
\\
\vspace{1mm}
(b) Real-world noisy labels.
\label{tab:noisy}
\end{table}


\begin{table}[!t]
\centering
\caption{Test set accuracy under imbalanced class distributions.}
\setlength\tabcolsep{6pt}
\centering
\begin{tabular}{cccccc}
\toprule
\multirow{2}{*}{Dataset} & \multirow{2}{*}{Method} & \multicolumn{4}{c}{Imbalance factor}  \\
                         &                          & 5 & 10 & 20 & 50  \\
\midrule
\multirow{3}{*}{CIFAR10} 
& Uniform & 92.2\% & 90.1\% & 87.1\% & 81.2\%   \\
& FSR & 93.4\% & 90.1\% & 87.2\% & 81.7\%   \\
& Ours  & \textbf{93.6\%} & \textbf{90.3\%} & \textbf{87.8\%} & \textbf{82.1\%}  \\
\midrule
\multirow{3}{*}{CIFAR100} 
& Uniform & 63.8\% & 31.2\% & 32.6\% & 34.9\%   \\
& FSR  & 69.8\% & 63.9\% & 56.8\% & 48.9\% \\
& Ours & \textbf{70.2\%} & \textbf{64.5\%} & \textbf{57.2\%} & \textbf{49.4\%}  \\
\bottomrule
\end{tabular}
\\
\vspace{1mm}
(a) Simulated imbalanced class distributions.
\vspace{2mm}

\setlength\tabcolsep{11pt}
\centering
\begin{tabular}{cccc}
\toprule
Dataset & Imbalance factor & Method & Accuracy \\
\midrule
\multirow{3}{*}{OCT} & \multirow{3}{*}{10}
& Uniform & 78.1\%    \\
& & FSR & 82.8\%    \\
& & Ours & \textbf{83.4\%}    \\
\bottomrule
\end{tabular}
\\
\vspace{1mm}
(b) Real-world imbalanced class distributions.
\vspace{-3mm}
\label{tab:imbal}
\end{table}

\setlength\tabcolsep{1pt}
\begin{table}[!t]
\centering
\caption{Test set accuracy under combined noise.}
\begin{tabular}{ccccccccc}
\toprule
Dataset & \# labels per class & Imbal. factor & Uniform & FSR & Ours\\
\midrule
\multirow{4}{*}{CIFAR10} 
& 10 & 5  &55.1\%  & 57.4\% & \textbf{58.0\%} \\ 
& 10 & 10 &54.2\% & 56.3\% & \textbf{56.9\%} \\ 
& 20 & 5  &62.4\%  & 62.8\% & \textbf{63.2\%} \\ 
& 20 & 10 &61.2\% & 62.1\% & \textbf{62.7\%} \\ 
\midrule
\multirow{4}{*}{CIFAR100} 
& 10 & 5  &29.5\%  & 30.4\% & \textbf{30.8\%} \\ 
& 10 & 10 &28.2\%  & 29.0\% & \textbf{29.7\%} \\ 
& 20 & 5  &34.7\%  & 35.0\% & \textbf{35.2\%} \\ 
& 20 & 10 &32.9\% & 33.4\% & \textbf{33.8\%} \\ 
\midrule
\multirow{2}{*}{Clothing} 
& N/A & 5  & 55.9\% & 59.8\% & \textbf{62.1\%} \\ 
& N/A & 10 & 52.6\% & 57.4\% & \textbf{59.8\%} \\ 
\midrule
\multirow{2}{*}{OCT} 
& 10 & N/A & 49.5\% & 57.4\% & \textbf{59.6\%} \\ 
& 20 & N/A & 53.0\% & 63.8\% & \textbf{65.2\%} \\ 
\bottomrule
\end{tabular}
\label{tab:combined}
\end{table}

\noindent\textbf{Results.}
Table~\ref{tab:noisy} shows that our method achieves better performance in the noisy label scenario.
The performance gain increases with the decreasing number of labeled samples, 
which demonstrates the effectiveness of our method in handling more label noise.
We also noticed that the gain on the Clothing dataset is larger than the gain on the CIFAR datasets.
The clearer differences between the classes of the CIFAR datasets enable FSR to easily select high-quality validation samples.
As a result, there is less room for additional improvements.
Unlike the CIFAR datasets, the Clothing dataset exhibits more subtle differences between its classes, making classification challenging. 
This type of task is known as fine-grained classification~\cite{lee2018cleannet}.
These subtle differences pose difficulties for the FSR method in selecting high-quality validation samples. 
In comparison, our method consistently improves the quality of validation samples and thus brings more performance gain.
In the imbalanced class distribution scenario, our method improves the classification accuracy on both the simulated and real-world datasets (Table~\ref{tab:imbal}).
For the retinal OCT images that are hard to classify correctly, our method still achieves better performance (Table~\ref{tab:imbal}(b)).
Table~\ref{tab:combined} shows that in scenarios where both noisy labels and imbalanced class distributions are present, our method achieves a larger performance gain.
The performance gain on the OCT dataset and the Clothing dataset reached around 2\%, exceeding that on the CIFAR datasets.
This further demonstrates that our method is more effective to improve validation samples in fine-grained classification, a task frequently encountered in real-world applications.
Due to the page limit, we only present the results with small numbers of labeled samples per class and small imbalance factors.
The full results are in supplemental material.

\subsection{Case Studies}
To demonstrate how {\sys} facilitates the human-AI collaboration in improving the reweighting results, we conducted two case studies.
They started from the output of the automatic reweighting method.
The case study results showed that users could further improve the accuracy by interactively providing a small amount of feedback.\looseness=-1

\begin{figure}[!b]
\centering
{
\vspace{-3mm}
\includegraphics[width=\linewidth]{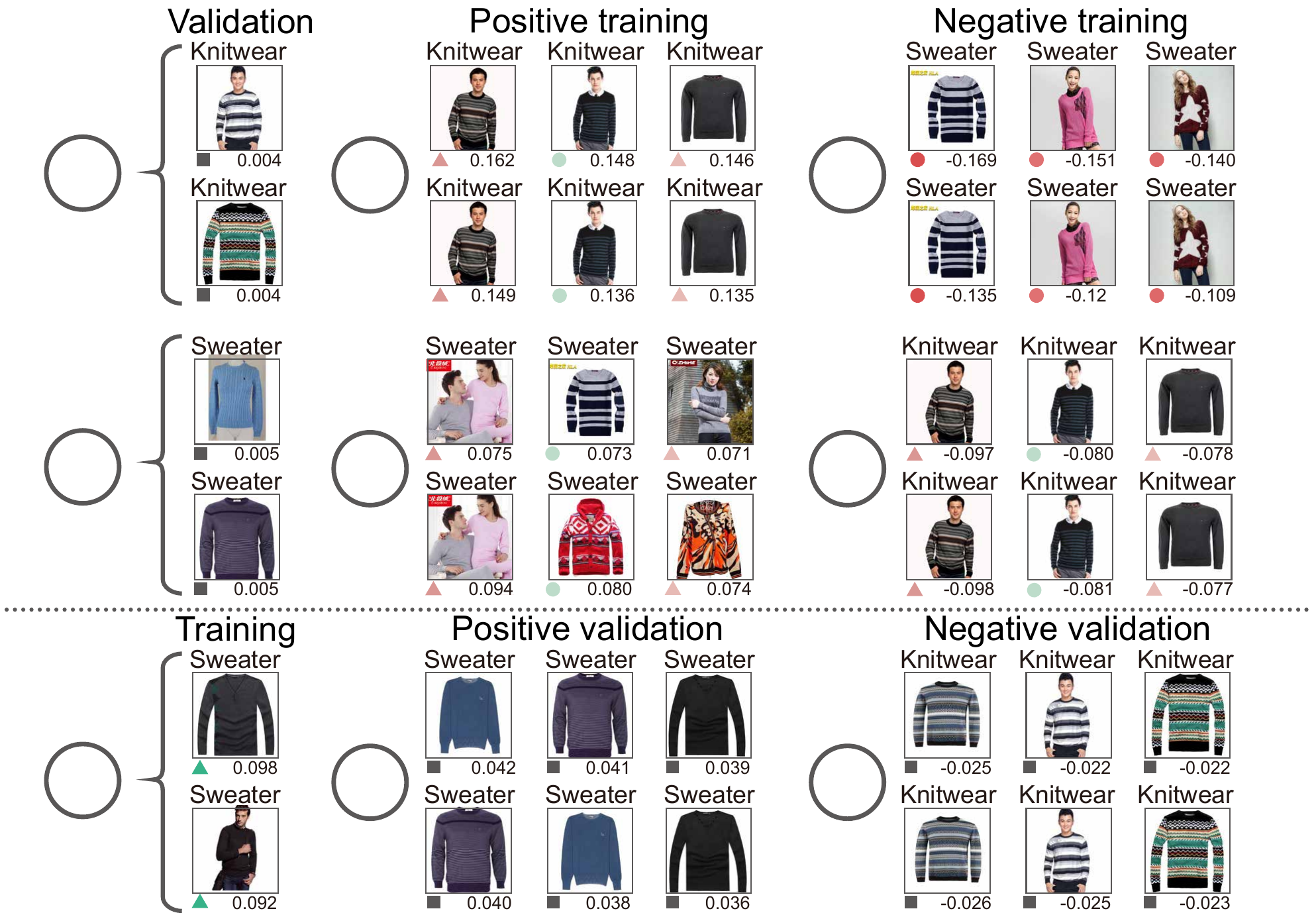}}
\put(-241,139){A}
\put(-187,139){A${}_\text{P}$}
\put(-95.5,139.5){A${}_\text{N}$}
 \put(-241,82.5){B}
\put(-186.5,82.5){B${}_\text{P}$}
\put(-95,83){B${}_\text{N}$}
\put(-240.5,23){C}
\put(-186.5,23){C${}_\text{P}$}
\put(-95,23){C${}_\text{N}$}
\vspace{-2mm}
\caption{Analyzing low-quality validation samples in clusters \V1 (A) and \V2 (B) and inconsistent training samples in cluster \S2 (C).\looseness=-1
}
\label{fig:clothsample}
\end{figure}

\subsubsection{Interactively Reweighting the Clothing Dataset}
In this case study, we collaborated with \E1 to interactively improve the reweighting results of the Clothing dataset.
This dataset, sourced from real-world data, contains many noisy labels.
The initial accuracy without human intervention was 71.9\% (Table~\ref{tab:noisy}(b)).

\noindent\textbf{Overview} (\R1). %
After co-clustering, there were 14 validation sample clusters and 35 training sample clusters.
Fig.~\ref{fig:teaser}(a) shows three validation sample clusters and six training sample clusters among them.
\E1 first looked at the validation sample clusters and noticed that in each cluster, there was no big difference in the $y$-positions of the validation samples. 
This is in line with the expectation that each validation sample should favorably contribute to the reweighting results.
However, the $y$-positions of the validation samples in clusters \V1 and \V2 are lower than the dotted line (the average value), indicating that their contributions are relatively low.
Thus, \E1 decided to examine them first.

\noindent\textbf{Identifying the low-quality samples} (\R1, \R2).
\E1 first selected all the 20 validation samples in clusters \V1 and \V2.
These validation samples and the training samples that were highly influenced by them were displayed in the sample view.
After examining the image content, he found that some validation samples in \V1 were actually ``sweater'' but mislabeled as ``knitwear'' (Fig.~\ref{fig:clothsample}A), while some samples in \V2 were actually ``knitwear'' but mislabeled as ``sweater'' (Fig.~\ref{fig:clothsample}B).
Similar mislabeling issues were also identified in their associated training samples.
For example, the samples in Fig.~\ref{fig:clothsample}A${}_\text{P}$ were ``sweater'' but mislabeled as ``knitwear.''
To explore more training samples that were highly influenced by the selected validation samples, \E1 turned to examine the training sample clusters.
He found that clusters \S1 and \S2 were highly influenced and contained the most inconsistent samples \glyph[1]{greentri} and \glyph[1]{redtri} (Fig.~\ref{fig:teaser}(a)).
\E1 selected these inconsistent samples to examine their content and that of their associated validation samples in the sample view.
Fig.~\ref{fig:clothsample}C shows two inconsistent samples \glyph[1]{greentri} with positive weights but low-confidence values in cluster \S2.
They were in fact ``knitwear'' but mislabeled as ``sweater,'' so as their most positively contributing validation samples (Fig.~\ref{fig:clothsample}C${}_\text{P}$).
The wrong labels of these validation samples explain the positive weights of low-quality samples in Fig.~\ref{fig:clothsample}C.
\E1 concluded that the label noise in \V1 and \V2 led to the inconsistencies in \S1 and \S2.

\begin{figure*}[b]
\centering
{\includegraphics[width=\linewidth]{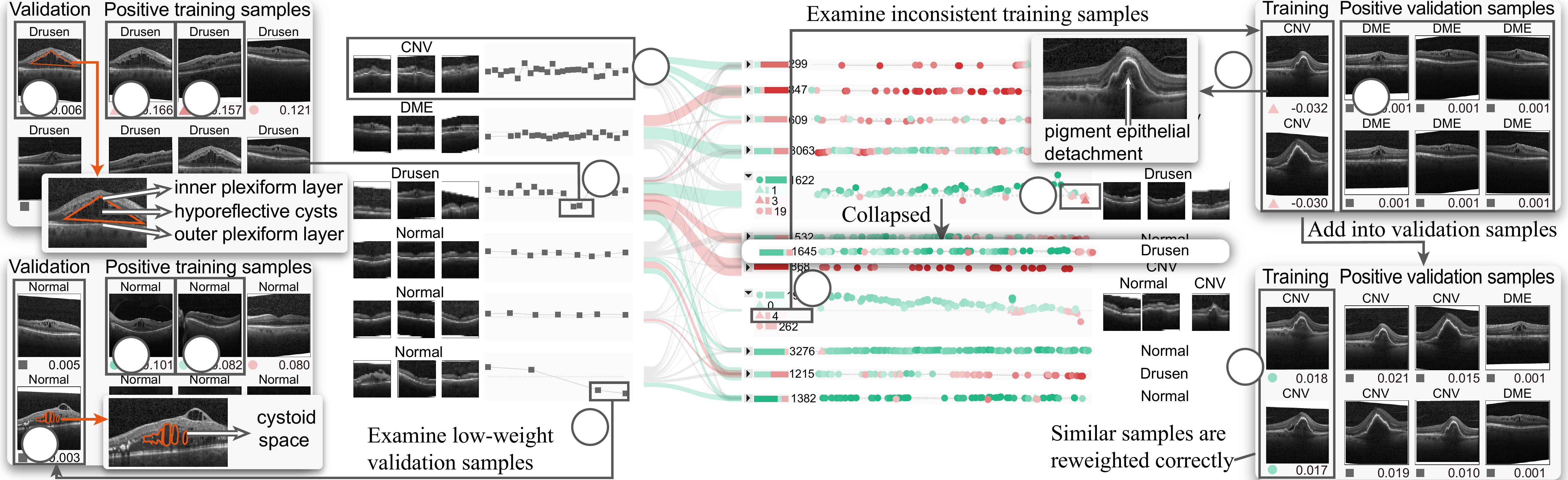}}
\put(-327,14.7){A}
\put(-510,8.6){\fontsize{6}{6}{A${}_1$}}
\put(-480,38.5){\fontsize{6}{6}{A${}_2$}}
\put(-456,38.5){\fontsize{6}{6}{A${}_3$}}
\put(-323,96){B}
\put(-509.5,123){\fontsize{6}{6}{B${}_1$}}
\put(-479.5,123){\fontsize{6}{6}{B${}_2$}}
\put(-456,123){\fontsize{6}{6}{B${}_3$}}
\put(-180,90.5){\fontsize{6}{6}{B${}_4$}}
\put(-306.5,133.5){D}
\put(-253,59.5){C}
\put(-110,34){E}
\put(-149,90){E}
\put(-114.5,132.5){\fontsize{6}{6}{C${}_1$}}
\put(-70,123){\fontsize{6}{6}{C${}_2$}}
\caption{Interactively improving the quality of validation samples in the OCT case study.}
\label{fig:medical-case}
\end{figure*}

\noindent\textbf{Improving the quality of validation samples} (\R3, \R4).
From these observations, \E1 decided to
1) correct the noisy labels in clusters \V1 and \V2,
2) increase the weights of the high-quality validation samples in \V1 and \V2, and
3) examine the inconsistent training samples \glyph[1]{greentri} and \glyph[1]{redtri} in clusters \S1 and \S2 and verify them as high/low-quality samples.
In total, he corrected the labels of 9 validation samples, increased the weights of 11 validation samples, and verified 49 training samples.
After the adjustments (Fig.~\ref{fig:teaser}(b)), the inconsistency in cluster \S2 was eased, and hence the cluster was collapsed automatically (Fig.~\ref{fig:teaser}\SP2).
However, as shown in the bar chart on the left of Fig.~\ref{fig:teaser}\SP1, there were more inconsistent samples \glyph[1]{redtri} with high-confidence values but negative weights in the training sample cluster \SP1. 
To understand the reason, \E1 compared the reweighting results before and after the adjustments in the ``diff'' mode. 
On the right of Fig.~\ref{fig:teaser}\SP1, he found many red lines.
This indicated that many high-confidence samples in \S1 had a large drop in their weights and thus resulted in more inconsistency.
Examining the image content of these samples, he found that they were samples of other classes but mislabeled as ``knitwear.''
These low-quality samples were not identified in the previous step due to their incorrect positive weights coinciding with their incorrect high-confidence values.
The adjustments he made shifted their weights to the negative side, revealing the inconsistency. 
E1 then verified these 12 samples as low-quality samples to reduce their confidence values.
As shown in Fig.~\ref{fig:teaser}$\text{S}_1''$, there are fewer inconsistent samples \glyph[1]{redtri}.
The adjustments were used to fine-tune the model, leading to an accuracy boost from 71.9\% to 72.7\%.

Similar mislabeling issues were observed in ``down coat,'' ``jacket,'' ``windbreaker,'' ``vest,'' and ``dress.''
\E1 corrected the labels of 13 validation samples and verified 71 training samples as high/low-quality samples.
Upon satisfaction, he obtained a set of better-quality validation samples and a set of verification of high/low-quality training samples, which were used to fine-tune the model.
The accuracy was boosted from $72.7\%$ to $75.4\%$.
In summary, \E1 further improved the model accuracy from 71.9\% to 75.4\% (+3.5\%) by correcting the labels of 22 validation samples and verifying 132 training samples as high/low-quality samples.
\looseness=-1

\subsubsection{Interactively Reweighting the OCT Dataset}
In this case study, we collaborated with \E3 to improve the reweighting results of the OCT dataset with added noisy labels (generated from 20 labeled samples per class), which was used in the aforementioned combined scenarios. 
This dataset is more complex because it contains both noisy labels and imbalanced class distributions.
\E3 has experience in developing models for diagnosing retinal edema.
We also invited \D1, an ophthalmology clinical doctor, to examine medical images and explain his judgments based on medical expertise.
The initial accuracy without human intervention was 65.2\% (the last row in Table~\ref{tab:combined}).

\noindent\textbf{Overview} (\R1). 
After co-clustering, there were 6 validation sample clusters and 11 training sample clusters.
\E3 noticed that the weights of validation samples in Fig.~\ref{fig:medical-case}A and Fig.~\ref{fig:medical-case}B are lower than other validation samples in the same clusters.
This indicated that these samples did not favorably contribute to the reweighting results.
He decided to analyze them first.

\noindent\textbf{Diagnosing \& Improving low-weight validation samples} (\R1, \R2, \R3).
As the two validation samples in Fig.~\ref{fig:medical-case}A (with the label ``Normal") had the lowest weight, \E3 examined them first.
By examining their image content (Fig.~\ref{fig:medical-case}A${}_1$) and the associated training samples, he found that these two validation samples 
and two of the training samples (Fig.~\ref{fig:medical-case}A${}_2$ and Fig.~\ref{fig:medical-case}A${}_3$) contained cystoid space (the dark chamber in the middle).
This makes them different from other ``Normal" samples.
He then consulted with doctor \D1, who confirmed that both the validation samples and the two training samples were ``DME'' but mislabeled as ``Normal.'' 
He explained 
that the cystoid space was a typical symptom of ``DME.''
However, the green circles \glyph[1]{green} indicated that these two training samples were wrongly assigned positive weights. 
He corrected the labels of these two validation samples and verified the two training samples as low-quality samples.
Next, \E3 analyzed the validation samples in Fig.~\ref{fig:medical-case}B in a similar way. 
\D1 confirmed that these two samples were ``DME'' but mislabeled as ``Drusen.''
The hyporeflective cysts between the inner plexiform layer and the outer plexiform layer form a triangle with white borders (Fig.~\ref{fig:medical-case}B${}_1$), which is the symptom of ``DME.''
After examining the highly influenced training samples, he found two inconsistent training samples \glyph[1]{redtri} with similar symptoms (Fig.~\ref{fig:medical-case}B${}_2$, B${}_3$) were also ``DME'' but mislabeled as ``Drusen.''
These training samples appeared in Fig.~\ref{fig:medical-case}B${}_4$, where he found more similar samples that were mislabeled as ``Drusen.''
\E3 corrected the labels of these two low-quality validation samples and verified seven training samples in Fig.~\ref{fig:medical-case}B${}_4$ as low-quality samples.
After these adjustments, the inconsistency in the training sample cluster ``Drusen" was reduced, and the cluster was collapsed automatically. The accuracy was improved from 65.2\% to 66.1\%.\looseness=-1

\noindent\textbf{Adding missing validation samples} (\R1, \R2, \R3).
After adjusting the validation samples, \E3 turned to examine the training samples. 
He noticed that there were four inconsistent samples \glyph[1]{redtri} with high-confidence values but negative weights in a cluster containing both ``Normal" and ``CNV" samples (Fig.~\ref{fig:medical-case}C).
He clicked the bar to examine their content and the contributing validation samples.
The image content (Fig.~\ref{fig:medical-case}C${}_1$) showed that they were high-quality samples of ``CNV.''
However, there were no validation samples of ``CNV'' that positively influenced them (Fig.~\ref{fig:medical-case}C${}_2$).
He then examined the validation samples in the validation sample cluster ``CNV'' (Fig.~\ref{fig:medical-case}D) and found that these samples were not representative enough for the ``CNV'' class.
According to doctor \D1, the ``CNV'' training samples in Fig.~\ref{fig:medical-case}C show serious pigment epithelial detachments adjacent to a small area of subretinal fluid (the sharp protrusion), which is one of the typical symptoms of ``CNV.''
However, none of the existing validation samples in ``CNV'' have this symptom.
This explained why these training samples received negative weights. 
To address this issue, \E3 added two of them into the validation set and updated the reweighting results. 
The other two training samples (Fig.~\ref{fig:medical-case}E) were then correctly assigned positive weights due to the added validation samples.
The model was fine-tuned with these added validation samples, and the accuracy was increased from 66.1\% to 68.1\%.
In summary, \E3 further improved the model accuracy from 65.2\% to 68.1\% (+2.9\%) by adding two validation samples, correcting the labels of four validation samples, and verifying nine training samples as high/low-quality samples.\looseness=-1

\subsection{Comparison with DataDebugger}
To demonstrate the effectiveness of {\sys}, we compared it with DataDebugger~\cite{xiang2019interactive}, which is the state-of-the-art visual analysis tool for interactively correcting label noise in training samples.
In DataDebugger, users need to examine samples and provide exact labels for them.
The provided labels are then propagated to other samples with a label correction algorithm.
However, since no additional information is considered when correcting the labels, more samples with exact labels are required to achieve acceptable performance.
In contrast, {\sys} enables users to analyze the reweighting relationships between validation samples and training samples.
This additional relationship information facilitates the correction of label noise in validation samples and the verification of high/low-quality training samples.
These corrections and verification help improve the reweighting results and boost the model performance.
In this experiment, we invited \E1-\E4 to use both tools and recorded their adjustments and time spent.
All the experts are familiar with both tools.
To reduce the potential bias resulting from the order in which the tools were used, \E1 and \E2 were asked to use DataDebugger first, while \E3 and \E4 were asked to use {\sys} first.
The Clothing dataset was employed because they are more familiar with it, and the initial accuracy was 71.9\% for both tools.
Table~\ref{tab:case-compare} shows the results from each expert.
It can be seen that to achieve comparable performance, {\sys} requires an average of 22.75 exact labels, which is much fewer than 297.25 labels that DataDebugger needs.
This is because DataDebugger requires the experts to label training samples and then propagate them to other samples.
In contrast, {\sys} only requires the experts to label validation samples, which are much fewer than training samples (140 compared to 36,000).
Although {\sys} requires additional verification of high/low-quality training samples, the average number of verified samples (156.25) is still less than the average number of exact labels required by DataDebugger (297.25).
Moreover, \E1 pointed out that providing verification is easier than providing exact labels, particularly with 14 classes in the Clothing dataset.
The verification effort is further simplified because high/low-quality samples are usually grouped together in {\sys}.
As illustrated in Fig.~\ref{fig:teaser}, the 18 inconsistent training samples with low weights in cluster \S1 can be collectively verified as low-quality samples with just one click.
In contrast, since their ground-truth labels are different (``knitwear,'' ``sweater,'' ``shawl,'' and ``underwear''), the experts have to re-label each sample separately, which takes more time.
On average, they spent 0.57 hours using Reweighter, which was shorter than the 1.49 hours they spent with DataDebugger to achieve comparable performance.

\setlength\tabcolsep{2.5pt}
\begin{table}[!ht]
\centering
\caption{
The numbers of provided exact labels, verification, and time comparison between DataDebugger and {\sys}.
}
{
\begin{tabular}{lccccc}
 \toprule
 Method  & Expert & \#  labels & \# verification & Time & Accuracy \\
  \midrule
  \multirow{4}{*}{DataDebugger} 
  & \E1   & 308 & 0  & 1.66h & 75.0\%   \\
  & \E2   & 293 & 0  & 1.50h & 74.7\%   \\
  & \E3   & 331 & 0  & 1.63h & 75.1\%   \\
  & \E4   & 257 & 0  & 1.17h & 74.6\%   \\
 \midrule
 \multirow{4}{*}{{\sys}} 
   & \E1 & 22 & 132  & 0.54h & 75.4\% \\
   & \E2 & 26 & 179  & 0.62h & 75.6\% \\
   & \E3 & 25 & 171  & 0.66h & 75.4\% \\
   & \E4 & 18 & 143  & 0.47h & 75.1\% \\
 \bottomrule
\end{tabular} 
}
\vspace{-3mm}
\label{tab:case-compare} 
\end{table}

\section{Expert Feedback and Discussion}
After the case studies, we conducted six semi-structured interviews with the four experts we collaborated with (\E1-\E4) and two newly invited ones (\E5 and \E6).
For the experts who were not involved in the case studies, we first introduced {\sys} and then presented the case studies.
Then the experts were asked to identify low-quality validation samples and make adjustments using the tool.
Finally, we discussed with the experts about the strengths and limitations of the tool.
The entire process lasted from 50 to 85 minutes.

The experts were generally positive about the usability of {\sys}.
A few limitations were also identified by the experts, from which we summarized several future research directions.\looseness=-1

\subsection{Usability}
\noindent\textbf{Adopting simple visual design}.
All the experts appreciated the simplicity of the visual design used in {\sys}.
They indicated that the cluster view was intuitive and easy to understand, facilitating the identification of low-quality validation samples.
\E4 commented, ``The visual encoding of validation samples is concisely explained in the legend.
Guided by the explanation, I can quickly identify the low-weight validation samples and select them in the cluster view.
Moreover, the bipartite graph provides a clear overview of their reweighting relationships, enabling me to conveniently trace the related training samples for more comprehensive examination.''
The experts also emphasized the usefulness of the sample view in identifying low-quality validation samples. 
Their prior experience with list-based visual designs made this view particularly accessible and streamlined their analysis.
For example, \E5 mentioned,
``During an analysis, after selecting a few inconsistent training samples, I found that some validation samples appeared repeatedly in the sample view.
More examination showed that these repetitively occurring samples were of low quality and required curation.''
He added that this simple but intuitive design made these recurring samples stand out.

\noindent\textbf{Providing an efficient way to diagnose quality issues}.
All the experts were able to efficiently identify low-quality validation samples and made appropriate  adjustments.
They agreed that {\sys} reduced their efforts in diagnosing quality issues of validation samples.
\E2 commented, ``The glyphs \glyph[1]{greentri}/\glyph[1]{redtri} are useful in highlighting inconsistent training samples for further examination.
These samples often reflect quality issues in validation samples.''
Meanwhile, \E5 highlighted the practicality of the ``diff'' mode, noting, ``It provides a convenient method for me to examine weight changes and evaluate the appropriateness of my adjustments.''
Interestingly, our observations revealed a divergence in expert strategies: a subset of experts (\E1, \E3, \E4, \E6) usually started their analysis from low-weight validation samples, while the others (\E2, \E5) preferred to start their analysis from inconsistent training samples.
This indicates that different analysis pipelines offered by our tool meet the diverse analytical preferences of individual experts.
After examining the quantity evaluation and detailed case studies, \E1 concluded that {{\sys} is effective at handling fine-grained classification, which aims to distinguish different classes that are closely related to each other (\eg, ``knitwear'' and ``sweater'' in the Clothing dataset).
Given the subtle differences in fine-grained classification, validation samples often have more label noise.
Our tool helps effectively identify and correct such label noise, and thus leads to a larger performance gain.
The experts were also satisfied with the performance gains on real-world datasets ranging from 1.4\% to 3.2\%.
\E6 commented, ``In real-world applications, even a performance gain close to 1\% is important and highly valued.
For example, a slight improvement in accuracy can enhance short video recommendations. 
This, in turn, leads to increased user engagement and extended interaction duration on the platform.''}

\noindent\textbf{Being easily integrated with different reweighting methods}.
Our work demonstrates how Reweighter improves the performance of FSR~\cite{zhang2021learning}.
Representing reweighting relationships as a bipartite graph offers the potential to integrate different reweighting techniques.
For example, \E6 indicated that our tool can be easily integrated with different reweighting methods by modeling the reweighting relationships as a bipartite graph.
He noted, ``The tool directly supports validation-sample-based reweighting methods by extracting the reweighting relationships between validation samples and training samples using the corresponding reweighting methods.''
As distribution-based methods~\cite{liu2015classification,liu2021label} do not include validation samples in the reweighting process, they cannot directly apply {\sys}.
In these situations, there is a need to understand their reweighting mechanism and then construct the corresponding bipartite graph.
For example, the method proposed by Liu~\etal~estimates the sample weights by assessing the embedding similarity between training samples and a set of selected trusted samples~\cite{liu2021label}.
These similarity relationships can be modeled as a bipartite graph, which enables the utilization of {\sys}.

\subsection{Limitations and Future Work}
\noindent\textbf{How to provide in-context recommendation}.
Currently, the experts started their analysis after identifying the samples of interest (\eg, low-quality validation samples).
To save exploration efforts, \E1, \E3, and \E5 expressed a preference for the tool to automatically recommend these samples to them.
A straightforward solution is to recommend low-weight validation samples and inconsistent training samples.
However, \E3 was not satisfied with this solution, ``I can conduct the same analysis by selecting validation samples with low y-positions and training samples with triangle shapes.
Such recommendations do not offer me additional benefits.''

A more favorable solution is to dynamically recommend the samples of interest based on users' previous adjustments.
As \E1 mentioned, ``If {\sys} can recommend more samples with similar quality issues after I have identified a few, it would greatly improve efficiency.
Such a method will streamline the analysis process  by maintaining a consistent context.''
The challenge lies in simultaneously considering the quality issues and context when making recommendations, which deserves further investigation.

\noindent\textbf{How to support online monitoring}.
After using our tool, \E6 suggested that it would be more helpful if {\sys} could analyze the reweighting relationships and results during the model training process,
``The reweighting relationships and results will gradually change as the model is trained over epochs.
If {\sys} supports tracking these dynamic changes, I can intervene and adjust validation samples to match current models better.''
This perspective was echoed by both \E1 and \E3.
After a thorough discussion, we reached a consensus that our tool could partially support this functionality by automatically updating the reweighting relationships during model training.
However, the analysis of these relationships and results across different epochs still presents challenges for model developers.
How to design an informative overview that effectively summarizes the changes over epochs deserves further exploration.
Moreover, in an online monitoring system, it is crucial to have an alert mechanism that can promptly report potential quality issues to model developers.
How to design an alert mechanism that can accurately report the issues and reduce false alarms requires more investigation.\looseness=-1

\noindent\textbf{How to extend to other tasks}.
Although our tool is evaluated with image classification, it directly supports the classification tasks of other types of data, including text, video, and chart~\cite{wu2022ai4vis,revision2011savva}.
This is because the extraction of reweighting relationships is not limited to specific types of data.
The experts also discussed the potential of applying our tool to other tasks, such as object detection and image segmentation.
The bipartite graph construction method, co-clustering algorithm, and visualization can be directly applied to other tasks because their design is inherently task-agnostic, focusing on reweighting relationships rather than specific task details.
However, the validation sample improvement method needs adaptation to address the complexities of object detection and image segmentation tasks.
Elaborating on this, \E6 highlighted the differences between the tasks: ``In classification, I only need to verify whether the labels are clean, which can be performed in batch mode.
However, object detection and image segmentation require additional feedback on the bounding boxes and segmentation masks, making batch annotation adjustment unfeasible.
Developing strategies to address the unique challenges of object detection and image segmentation in the context of validation sample improvement remains a promising avenue for future research.''

\section{Conclusion}
We have developed {\sys}, a visual analysis tool for generating better reweighting results of training samples.
The key is to model the reweighting relationships between validation samples and training samples as a bipartite graph.
Based on this graph, we develop a validation sample improvement method and a co-cluster-based bipartite visualization.
They are tightly integrated together to support an interactive sample reweighting process,  
where the user adjustments are converted to the constraints of the validation sample improvement method. 
This process interactively improves the validation samples and hence generates better reweighting results.
A quantitative evaluation and two case studies are conducted to demonstrate the effectiveness and usefulness of {\sys} in improving sample reweighting methods.
}

\ifCLASSOPTIONcompsoc
  \section*{Acknowledgments}
\else
  \section*{Acknowledgment}
\fi
This research is supported by the National Natural Science Foundation of China (No.s U21A20469, 61936002), the National Key R\&D Program of China (No. 2020YFB2104100), grants from the Institute Guo Qiang, THUIBCS, and BLBCI.
The authors would like to thank Yuxing Zhou for his efforts in the validation sample weight adjustment algorithm, Dr. Changjian Chen and Yusong Zhu for their engaging discussion about the co-clustering algorithm, and Lanxi Xiao, Jing Wang, and Hanjie Yu for their valuable comments on the visualization design. 

\ifCLASSOPTIONcaptionsoff
  \newpage
\fi

\bibliographystyle{IEEEtran}
\bibliography{template}
\vspace{-12mm}

\begin{IEEEbiography}
[{\includegraphics[width=1in,height=1.25in,clip,keepaspectratio]{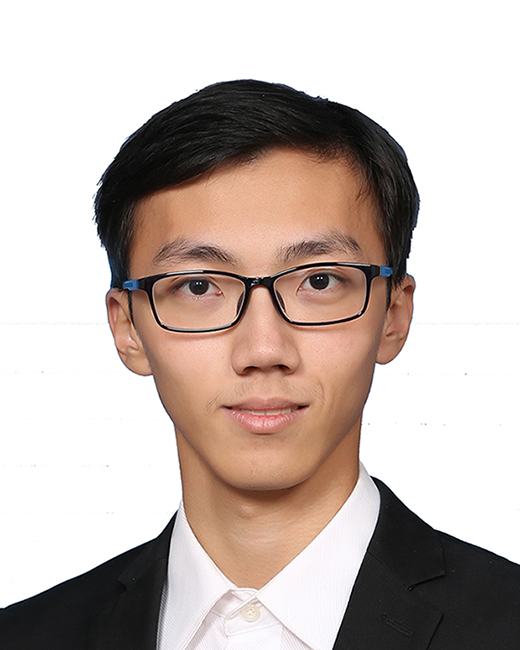}}]
{Weikai Yang} is a Ph.D. candidate at Tsinghua University. His research interests include visual text analytics and interactive machine learning. He received a B.S. degree from Tsinghua University.
\end{IEEEbiography}

\begin{IEEEbiography}
[{\includegraphics[width=1in,height=1.25in,clip,keepaspectratio]{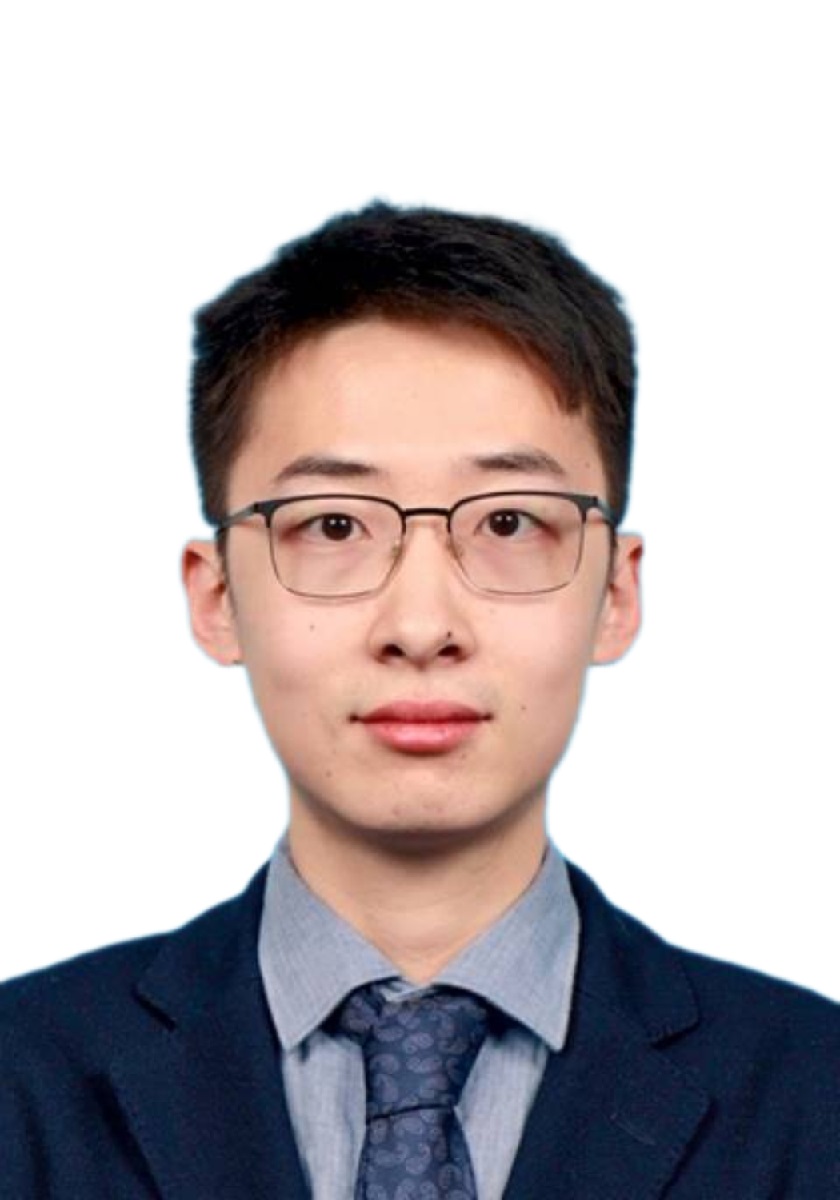}}]
{Yukai Guo} is a Ph.D. student at Tsinghua University. His research interests include interactive machine learning and model evaluation. He received a B.S. degree from Tsinghua University.
\end{IEEEbiography}

\begin{IEEEbiography}
[{\includegraphics[width=1in,height=1.25in,clip,keepaspectratio]{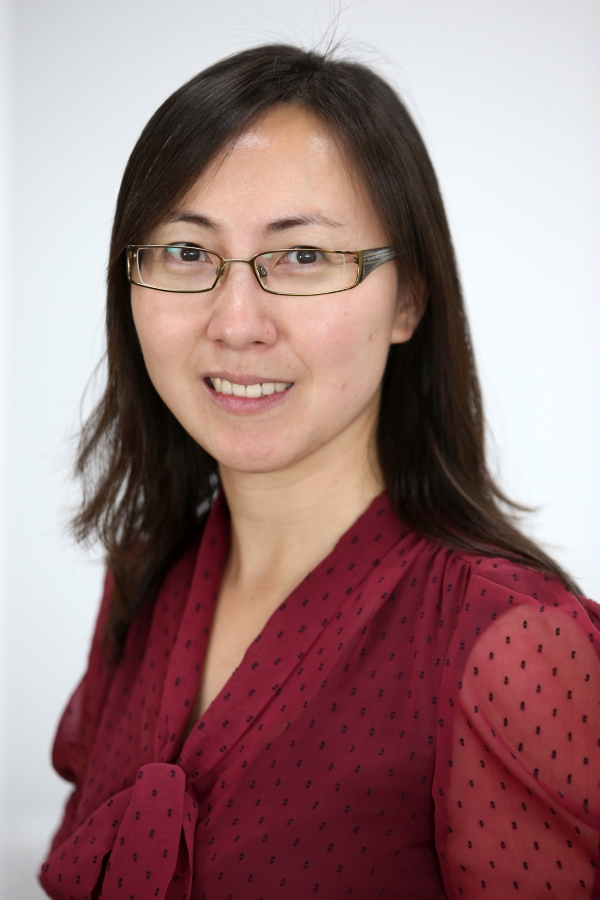}}]{Jing Wu} is a lecturer in computer science and informatics at Cardiff University, UK. Her research interests are in computer vision and graphics including image-based 3D reconstruction, face recognition, machine learning and visual analytics. She received BSc and MSc from Nanjing University, and Ph.D. from the University of York, UK. She serves as a PC member in CGVC, BMVC, etc., and is an active reviewer for journals including Pattern Recognition, Computer Graphics Forum, etc.
\end{IEEEbiography}

\begin{IEEEbiography}
[{\includegraphics[width=1in,height=1.25in,clip,keepaspectratio]{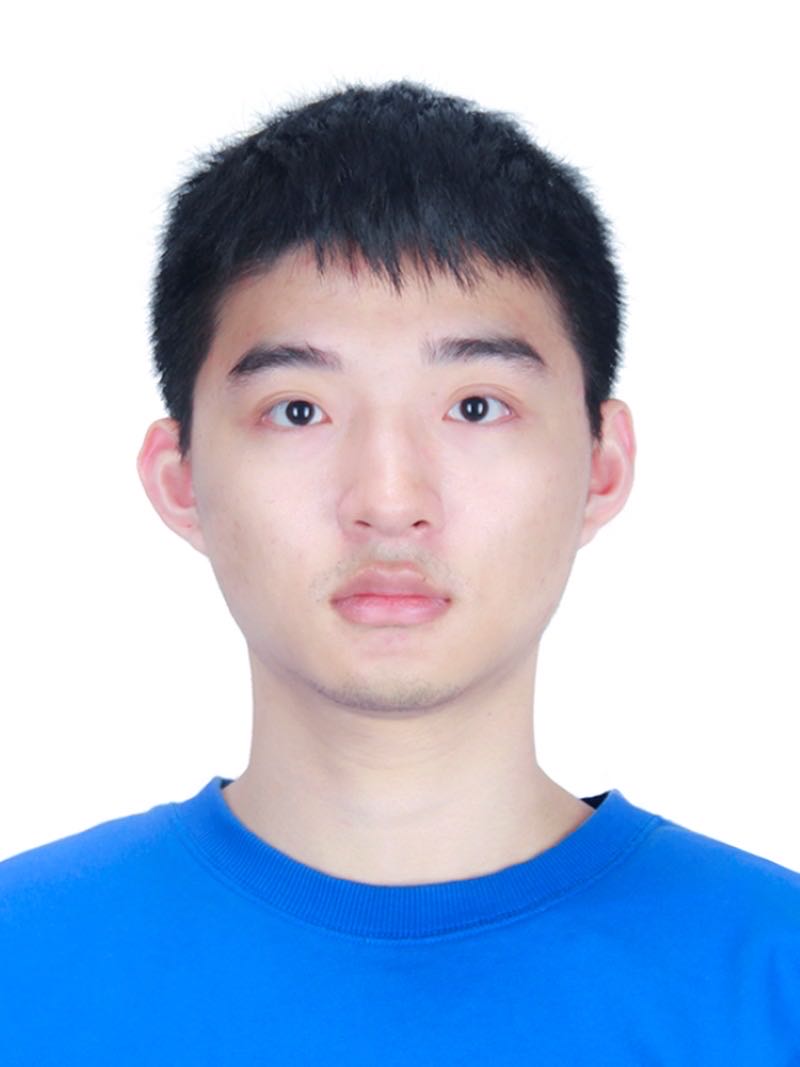}}]
{Zheng Wang} is currently working toward the graduate degree at Tsinghua University.
\end{IEEEbiography}

\begin{IEEEbiography}
[{\includegraphics[width=1in,height=1.25in,clip,keepaspectratio]{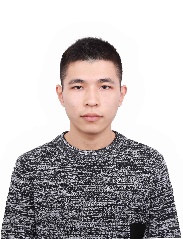}}]{Lan-Zhe Guo} received the BSc degree in 2017. He is currently working toward the Ph.D. degree in the National Key Laboratory for Novel Software Technology at Nanjing University, China. His research interests include semi-supervised learning, label noise learning, distribution shift learning. He is/was a PC member of top conferences such as AAAI, IJCAI, and a reviewer of journals like TKDE.
\end{IEEEbiography}

\begin{IEEEbiography}
[{\includegraphics[width=1in,height=1.25in,clip,keepaspectratio]{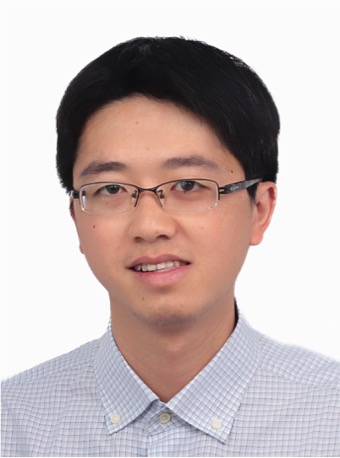}}]{{Yu-Feng Li}} is an associate professor in Nanjing University. His research interests include weakly supervised learning, statistical learning and optimization. He has received outstanding doctoral dissertation award from China Computer Federation (CCF) and Microsoft Fellowship Award. He is/was served as an editorial board member of machine learning journal special issues, co-chair of ACML18 workshop and ACML19 tutorial, and a senior PC member of top-tier conferences such as IJCAI’19/17/15, AAAI’19.
\end{IEEEbiography}

\begin{IEEEbiography}
[{\includegraphics[width=1in,height=1.25in,clip,keepaspectratio]{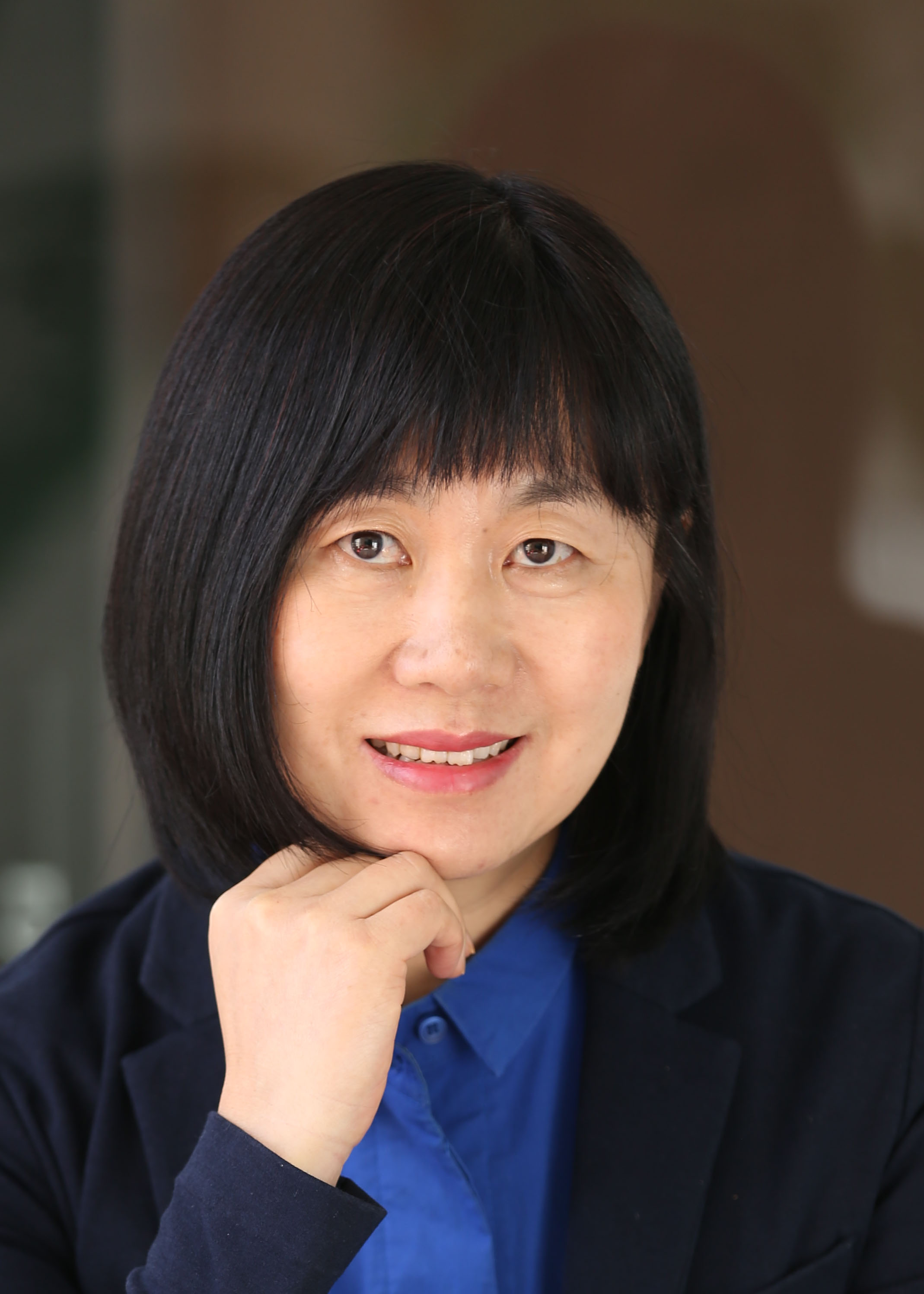}}]
{Shixia Liu}
is a professor at Tsinghua University. Her research interests include visual text analytics, visual social analytics, interactive machine learning, and text mining. She worked as a research staff member at IBM China Research Lab and a lead researcher at Microsoft Research Asia.
She received a B.S. and M.S. from Harbin Institute of Technology, a Ph.D. from Tsinghua University.
She is a fellow of IEEE and an associate editor-in-chief of IEEE Trans. Vis. Comput. Graph.
\end{IEEEbiography}

\end{document}